\documentclass[sigconf]{acmart}
\usepackage{multirow}
\usepackage{algorithm,algorithmic}

\usepackage{enumitem}
\setlist{topsep=0pt, leftmargin=*}
\AtBeginDocument{%
  \providecommand\BibTeX{{%
    \normalfont B\kern-0.5em{\scshape i\kern-0.25em b}\kern-0.8em\TeX}}}

\setcopyright{acmcopyright}
\copyrightyear{2018}
\acmYear{2018}
\acmDOI{10.1145/1122445.1122456}

\acmConference[Woodstock '18]{Woodstock '18: ACM Symposium on Neural
  Gaze Detection}{June 03--05, 2018}{Woodstock, NY}
\acmBooktitle{Woodstock '18: ACM Symposium on Neural Gaze Detection,
  June 03--05, 2018, Woodstock, NY}
\acmPrice{15.00}
\acmISBN{978-1-4503-XXXX-X/18/06}

\begin{document}

\title{Improving Multi-Interest Network with Stable Learning}


\author{Zhaocheng Liu}
\email{lio.h.zen@gmail.com}
\affiliation{
    \institution{Kuaishou Technology}
    \country{China}
}

\author{Yingtao	Luo}
\email{yl3851@uw.edu}
\affiliation{
    \institution{University of Washington}
    \country{United States}
}

\author{Di Zeng}
\email{zengdi19960922@163.com}
\affiliation{
    \institution{Kuaishou Technology}
    \country{China}
}

\author{Qiang Liu}
\email{qiang.liu@nlpr.ia.ac.cn}
\affiliation{
    \institution{Institute of Automation, Chinese Academy of Sciences}
    \country{China}
}

\author{Daqing Chang}
\email{changdaqing0906@126.com}
\affiliation{
    \institution{Kuaishou Technology}
    \country{China}
}

\author{Dongying Kong}
\email{kongdongying@kuaishou.com}
\affiliation{
    \institution{Kuaishou Technology}
    \country{China}
}

\author{Zhi	Chen}
\email{chenzhi07@kuaishou.com}
\affiliation{
    \institution{Kuaishou Technology}
    \country{China}
}







\renewcommand{\shortauthors}{Trovato and Tobin, et al.}

\begin{abstract}
Modeling users' dynamic preferences from historical behaviors lies at the core of modern recommender systems.
Due to the diverse nature of user interests, recent advances propose the multi-interest networks to encode historical behaviors into multiple interest vectors.
In real scenarios, the corresponding items of captured interests are usually retrieved together to get exposure and collected into training data, which produces dependencies among interests.
Unfortunately, multi-interest networks may incorrectly concentrate on subtle dependencies among captured interests.
Misled by these dependencies, the spurious correlations between \emph{irrelevant interests} and targets are captured, resulting in the instability of prediction results when training and test distributions do not match.
In this paper, we introduce the widely used Hilbert-Schmidt Independence Criterion (HSIC) to measure the degree of independence among captured interests and empirically show that the continuous increase of HSIC may harm model performance.
Based on this, we propose a novel multi-interest network, named DEep Stable Multi-Interest Learning (DESMIL), which tries to eliminate the influence of subtle dependencies among captured interests via learning weights for training samples and make model concentrate more on underlying true causation.
We conduct extensive experiments on public recommendation datasets, a large-scale industrial dataset and the synthetic datasets which simulate the out-of-distribution data.
Experimental results demonstrate that our proposed DESMIL outperforms state-of-the-art models by a significant margin.
Besides, we also conduct comprehensive model analysis to reveal the reason why DESMIL works to a certain extent.
\end{abstract}

\ccsdesc[500]{Information systems~Recommender systems}
\ccsdesc[500]{Information systems~Personalization}
\ccsdesc[500]{Information systems~Data mining}

\keywords{sequential recommendation, multi-interest, out-of-distribution, stable learning}


\maketitle

\section{Introduction}
Sequential recommender systems aim to predict the next item(s) that a user might be interested in based on historical interactions.
In the background of information explosion, they have become vital to alleviate the information overload problem and enhance user experiences.
Given historical behaviors, accurately characterizing and representing users' dynamic preferences is the core concern of research in sequential recommendation.
Traditional methods \cite{rendle2010factorizing,hidasi2016general} assume that the next action is conditioned on only the previous action (or previous few).
They adopt Markov chain and matrix factorization to capture short-range item transitions.
Due to the rapid development of deep learning, various deep neural networks are exploited to model the complex high-order sequential dependencies, including recurrent neural networks \cite{yu2016dynamic,hidasi2015session}, convolutional neural networks \cite{tang2018personalized} and attention mechanism-based networks \cite{liu2018stamp,wang2019towards,kang2018self,sun2019bert4rec,luo2021stan}.
With a user's historical behaviors, these deep learning-based approaches usually generate an overall embedding as user representation.
More recent advances \cite{li2019multi,cen2020controllable} argue that a unified user embedding is hard to encode the different aspects of the user’s interests.
Therefore, they propose the multi-interest networks which represent one user with multiple vectors to capture the user’s multiple interests and significantly outperform previous models.

\begin{figure}
    \centering
    \includegraphics[width=0.8\linewidth]{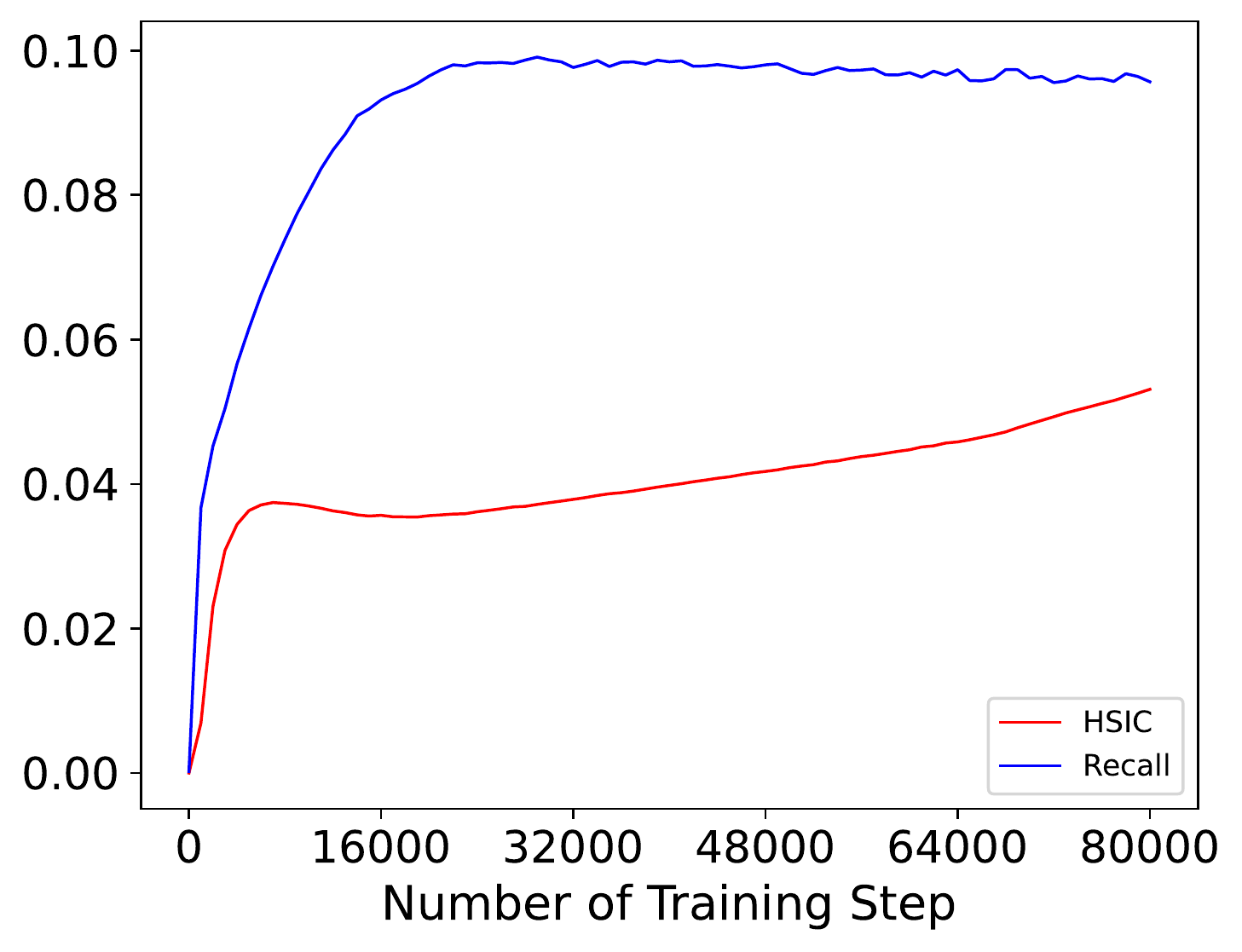}
    \caption{The curves of HSIC on the training set and Recall@50 on the validation set when training ComiRec on the Book dataset. In the early stage, both HSIC and recall increase rapidly. With the convergence of recall, HSIC further increases, which may harm the model performance.}
    \label{fig:baseline_hsic_recall}
\end{figure}

Despite the effectiveness of multi-interest networks, there are some challenges demanding further explorations.
A vital challenge is, multi-interest networks may concentrate more and more on the subtle dependencies among captured interests, which harms the generalization ability.
In this paper, we introduce the widely used Hilbert-Schmidt Independence Criterion (HSIC) \cite{gretton2007kernel,gretton2005measuring} to measure the degree of independence among captured interests. 
Then, we trace the change of HSIC on the training set and recall on the validation set when training the state-of-the-art multi-interest networks, i.e., ComiRec \cite{cen2020controllable}.
Taking Figure \ref{fig:baseline_hsic_recall} for example, due to the random initialization of model parameters, the value of HSIC and recall are nearly zero at first step, and they rapidly rise along with the increase of training steps.
For ComiRec, after 20000 steps, the value of HSIC (i.e., the red curve) begins to increase slowly while the value of recall (i.e., the blue curve) stops rising and even begins to decline.
To some extent, for ComiRec, this example reveals that excessive dependencies among interests may harm the model performance during inference stage, and limit the model's generalization ability.

\begin{figure}
    \centering
    \includegraphics[width=1.0\linewidth]{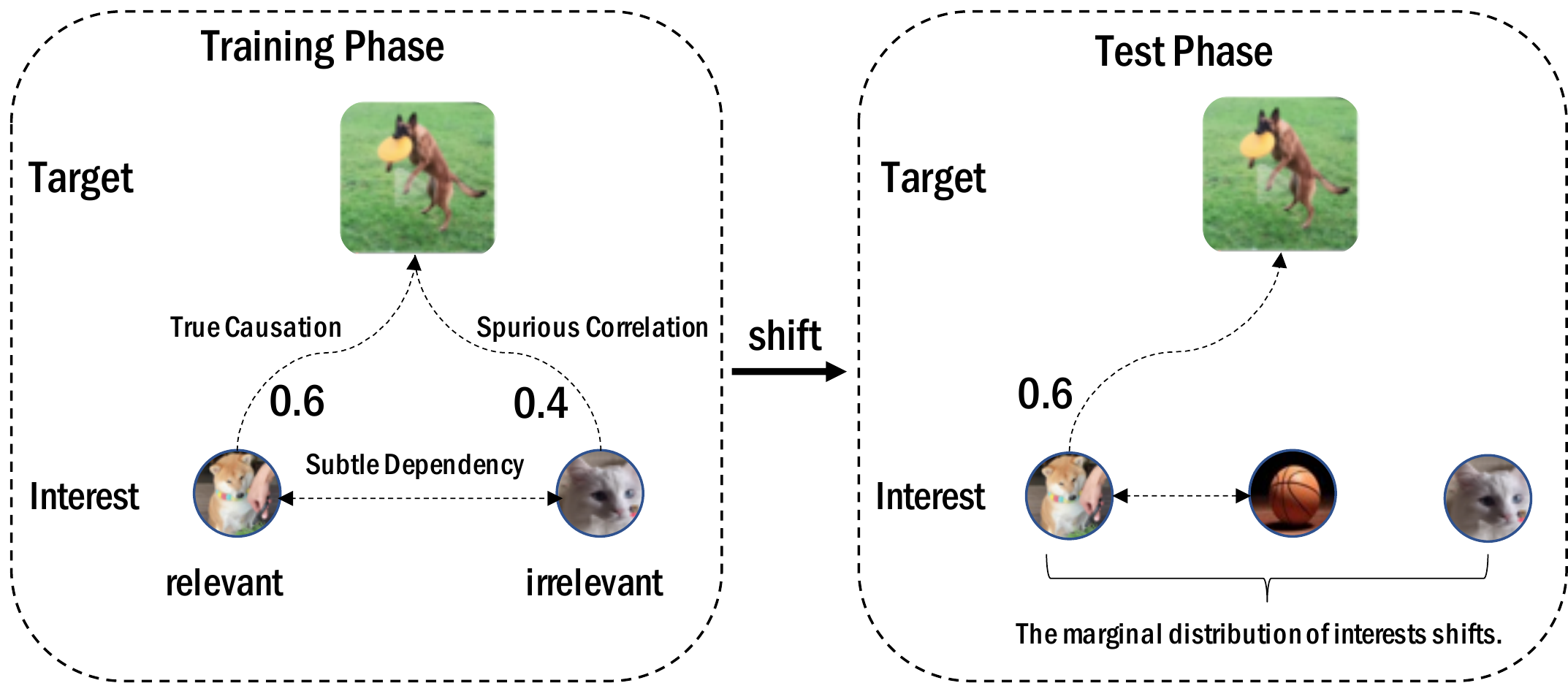}
    \caption{In training phase, due to interests of users and exposure in recommender systems during data collection, the user interactions may always contain dog and cat together. Misled by the subtle dependency, multi-interest networks may partially attribute to spurious correlation instead of focusing on true causation. However, in test phase, the subtle dependency between dog and cat is broken due to the rapid changing of recommender systems and the temporal evolution of user interests. Therefore, model fails to make trustworthy predictions.}
    \label{fig:intro_example}
\end{figure}

In real scenarios, as illustrated in Figure \ref{fig:intro_example}, we dissect above problem in recommender systems as follows.
Due to the existence of diverse interests of users and the noisy nature of implicit feedback \cite{o2006detecting,zhang2021causerec,wang2021denoising}, multi-interest networks can capture various interests, even including \emph{irrelevant interests} (i.e., the interests that are irrelevant to a given target item).
As prior work \cite{ma2018entire,yuan2019improving,chen2020esam} points out the sample selection bias problem, the corresponding items of captured interests are usually retrieved together to get exposure and collected into training data.
Therefore, there exists dependencies among captured interests.
Multi-interest models may be misled by these subtle dependencies, resulting in the spurious correlations between \emph{irrelevant interests} and target items are captured.
However, due to the rapid changing of recommender systems and the temporal evolution of user interest, the marginal distribution of captured interests shifts from training
phase to test phase, which breaks the subtle dependencies among multiple interests.
In such cases, the captured spurious correlations may be invalid in test phase, results in the drop of models’ performance, known as Out-Of-Distribution (OOD) generalization problem \cite{shen2021towards}.

To alleviate such OOD generalization problem, we aim to find a way to avoid multi-interest networks capturing the spurious correlations between \emph{irrelevant interests} and targets at all possible.
As it's hard to distinguish \emph{irrelevant interests} from captured interests, we turn to directly eliminating the dependencies among captured interests.
We propose a novel multi-interest network, named DEep Stable Multi-Interest Learning (DESMIL).
Inspired by sample reweighting techniques \cite{kuang2020stable,zhang2021deep}, the interest decorrelation regularizer in DESMIL aims to estimate a weight for each sample such that captured interests are decorrelated on the weighted training data.
Specifically, in training phase, given a specified sample, DESMIL mixes every captured interest's representation with sample weight, and adopts HSIC as independence testing statistics to measure any pair of such mixed representation.
DESMIL minimizes such statistics via finding optimal training sample weights while keeping the trainable parameters of multi-interest networks fixed.
Our design can ensure that the sample weight is reduced when the dependencies among captured interests' representation is strong.
Meanwhile, same sample weights are exploited to weight the classic sampled softmax loss and when optimizing such weighted loss, the sample weights are all fixed.
Therefore, the proposed DESMIL can make multi-interest networks concentrate more on the training samples whose dependencies between multiple interests are weak.

We conduct extensive experiments to verify the effectiveness of the proposed DESMIL on both public recommendation datasets and large-scale industrial dataset.
In particular, to validate the OOD performance of different models, we changes the splitting method of datasets to simulate different covariate shifts.
Experimental results demonstrate that our proposed DESMIL outperforms state-of-the-art models by a significant margin.
In order to reveal the reason why DESMIL works to a certain extent, we also conduct comprehensive model analysis including visualizing the probability distribution of the training sample weights learned by DESMIL and the change of Recall and HSIC of DESMIL through training steps.

To summarize, the main contributions of this paper are:
\begin{itemize}
\item We empirically show that excessive dependencies among interests may harm the model performance, and point out a promising way to solve the OOD generalization problem for multi-interest networks is to eliminate subtle dependencies among captured interests.
\item We propose a novel multi-interest network which eliminates dependencies between captured interests via learning weights for training samples.
\item Extensive experiments have been conducted and the effectiveness and OOD performance of the proposed DESMIL is fully verified.
\end{itemize}

\section{Related Work}
In this section, we review some works on sequential recommendation, deep multi-interest models and stable learning.
\subsection{Sequential Recommendation}
How to effectively extract user interests from users' historical behaviors is a critical problem in sequential recommendation \cite{fang2020deep}, which is a major task in recommender systems.
In some traditional models \cite{rendle2010factorizing,he2017translation,he2016fusing,hidasi2016general}, Markov chain and matrix factorization are widely used to model users' historical behaviors.
Among them, the most representative model is FPMC \cite{rendle2010factorizing}, which proposes to use a personalized Markov chain for capturing each user's behaviors, and train the model with a factorization model for capturing collaborative information.

With the rapid development of deep learning in past years, various deep neural networks such as recurrent neural network \cite{hidasi2015session,liu2016predicting,yu2016dynamic}, convolutional neural networks \cite{tang2018personalized,wang2019towards} and attention-based networks \cite{kang2018self, luo2021stan,liu2018stamp,sun2019bert4rec,li2020time} have been successfully exploited in the design of deep sequential recommendation model.
Generally, the user historical information is represented by the output of deep neural networks.
Thanks to the strong capacity of deep models, the recommendation accuracy has been greatly improved by these models \cite{fang2020deep}.
Recently, contrastive learning has been applied in sequential recommendation \cite{zhou2020s3,xie2020contrastive,liu2021contrastive}, for dealing with sparsity and noise in data.
Moreover, CauseRec \cite{zhang2021causerec} proposes to generate out-of-distribution counterfactual samples, and both model original samples and counterfactual samples with contrastive loss.

\subsection{Deep Multi-interest Models}

In real recommendation scenario, a user may have multiple interests in most time.  
Intuitively, the recommendation diversity is also a helpful property for user experience.
However, an overall user preference representation as in most works above can hardly grasp the diversity essence of user interests \cite{liu2019single}.
Starting from this, there are works \cite{zhou2021contrastive,chen2020improving,ma2019learning,li2019multi,cen2020controllable, tan2021sparse} studying how to effectively extract user's multiple interests in sequential recommendation with multiple user representations.


MIND \cite{li2019multi} proposes a multi-interest extractor layer based on the dynamic routing mechanism \cite{sabour2017dynamic, hinton2018matrix, hinton2011transforming}. 
As the procedure of dynamic routing can be seen as soft-clustering, the user's historical behaviors can be grouped into different clusters. 
Meanwhile, a label-aware attention mechanism is proposed to effectively aggregate the multiple user preference representations in training.
Besides, Cen et al. \cite{cen2020controllable} proposes a controllable multi-interest Framework called ComiRec. 
In ComiRec, both dynamic routing and self-attentive models can be adopted to extract multiple user interests. 
Furthermore, ComiRec shows that a controllable aggregation module balancing the accuracy and diversity is beneficial.
Lately, instead of implicitly generating user's multiple interests by clustering the user behaviors, SINE \cite{tan2021sparse} directly maintains a pool of conceptual prototypes to represent the all set of user's potential interests. 
Then a self-attention mechanism is used to decide which prototypes are activated as the user's multiple interests. 



\subsection{Stable Learning}

The out-of-distribution problem \cite{shen2021towards} is a common challenge in real-world scenarios, and stable learning has become a successful way to deal with this recently.
Stable learning aims to learn a stable predictive model that achieves uniformly good performance on any unknown test data \cite{kuang2018stable}.
To achieve this goal, the framework of most stable learning works can be divided into two steps: sample weight learning and weighted training.
Specifically, samples weights are learned to decorrelate features in training data, and then weighted training is conducted to train models on weighted feature distribution, which is approach to independent identically feature distribution.
Along this strand, various decorrelation methods\cite{shen2020stable, kuang2018stable, kuang2020stable, kuang2021balance} have been proposed to learn sample weights and train linear stable models.
Moreover, StableNet \cite{zhang2021deep} proposes to adopt random Fourier features to eliminate non-linear dependencies among features in convolutional neural networks.
And StableGNN \cite{fan2021generalizing} proposes to decorrelate features in graph neural networks.
Lately, Xu et al.\cite{xu2021stable} theoretically proves that the stability of least square regression and binary classification can be guaranteed with mutually independence of feature variables under mild conditions.

\begin{figure*}
    \centering
    \includegraphics[width=0.7\linewidth]{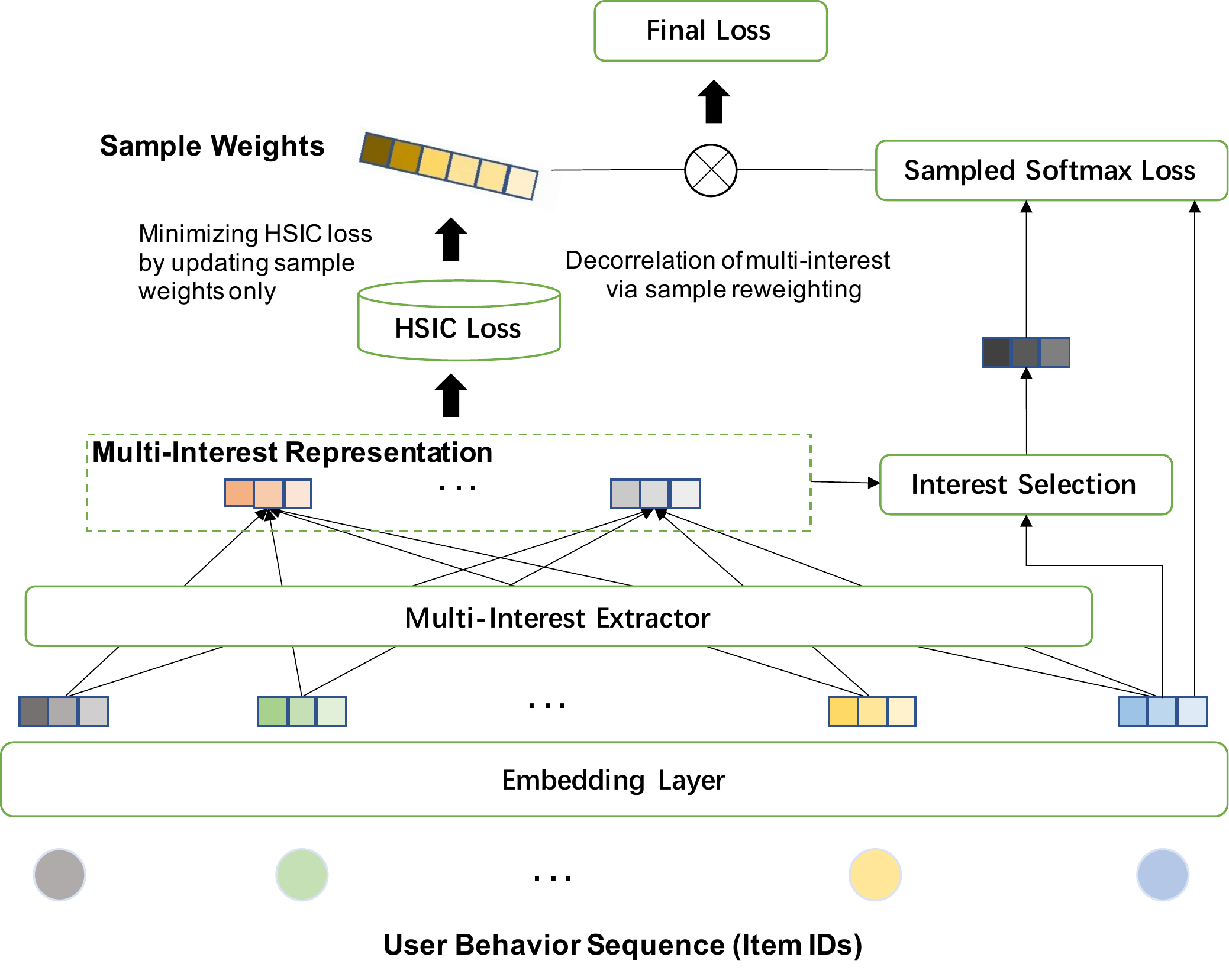}
    \caption{The overview of the proposed DESMIL. The input sequence is first embedded into dense representation to extract latent multi-interests. A HSIC loss is calculated based on multi-interest representations and optimized via sample weighting. The sample weights are then multiplied to the Softmax loss for final model optimization.}
    \label{fig:overview}
\end{figure*}
\section{METHODOLOGY}
In this section, we formulate the problem and introduce the proposed DESMIL in detail, and the overview of DESMIL is illustrated in Figure \ref{fig:overview}.
To be noted, in Section \ref{section:DependenciesAmongCapturedInterests}, we introduce HSIC as independence testing statistics and empirically show the relationship between HSIC and model performance in training phase.

\subsection{Problem Formulation}
In the setting of sequential recommendation, assume we have a set of users $\mathcal{U} = \{u_1, u_2,\cdots, u_{|\mathcal{U}|}\}$ and a universe of items $\mathcal{I} = \{i_1, i_2,\cdots, i_{|\mathcal{I}|}\}$.
For each user $u$, we have a sequence of historical behaviors $\mathcal{S}^u = (\mathcal{S}^u_1,\cdots,\mathcal{S}^u_{|\mathcal{S}^u|})$, where $\mathcal{S}^u_t \in \mathcal{I}$ and the index $t$ of $\mathcal{S}^u_t$ denotes the order of a specified behavior occurs in $\mathcal{S}^u$.
In training phase, for user $u$ at step $t$, the model's input and the expected output can be thought as $(\mathcal{S}^u_1,\cdots,\mathcal{S}^u_{t})$ and $\mathcal{S}^u_{t+1}$ respectively.
Given all users’ sequences $\mathcal{S}$, the goal of sequential recommendation is to recommend each user a list of items that maximize her/his future needs.
Besides, as each item is relevant to interests of user, the proposed DESMIL aims to capture the representation vectors of user interest.
We use $c$ to denote the number of such representation vectors.
\subsection{Embedding Layer}
We adopt the widely-used embedding technique to embed id features into into low-dimensional dense vectors.
Specifically, given the input sequence $(\mathcal{S}^u_1,\cdots,\mathcal{S}^u_{t})$, we create an embedding matrix $\mathbf{V} \in \mathbb{R}^{\mathcal{|I|}\times d}$ where $d$ is the number of latent dimensions, and retrieve the input embedding matrix by applying the embedding look-up operation.
Besides, to make the proposed DESMIL be aware of the positions of historical items, we inject the corresponding trainable position embedding matrix \cite{vaswani2017attention,kang2018self} $\mathbf{P} \in \mathbb{R}^{t\times d}$ into the input embedding matrix.
The final input embedding matrix $\mathbf{E} \in \mathbb{R}^{t\times d}$ can be formulated as
\begin{equation}
    \mathbf{E} = \left [ \begin{matrix}
    \mathbf{V}_{\mathcal{S}^u_1} + \mathbf{P}_1 \\
    \vdots \\
    \mathbf{V}_{\mathcal{S}^u_{t}} + \mathbf{P}_{t} \\
    \end{matrix} \right ].
\end{equation}
\subsection{Multi-Interest Extractor}
The multi-interest extractor is exploited to generate multiple representation vectors to capture diverse interests of users.
Prior multi-interest networks \cite{li2019multi,cen2020controllable} implements the multi-interest extractor via the dynamic routing mechanism \cite{sabour2017dynamic} or the attention mechanism \cite{lin2017structured}.
Empirically, as shown in prior work \cite{cen2020controllable,zhang2021causerec}, self-attention based multi-interest networks show the strong ability to capture user interests and get comparable results with the dynamic routing based methods.
Therefore, we adopt the self-attention mechanism to obtain an attention matrix $\textbf{A} \in \mathbb{R}^{c\times t}$ as
\begin{equation}
    \mathbf{A} = \textnormal{softmax}(\mathbf{W}_2\tanh(\mathbf{W}_1\mathbf{E}^\top)),
\end{equation}
where $\mathbf{W}_1 \in \mathbb{R}^{\hat{d}\times d}$ and $\mathbf{W}_2 \in \mathbb{R}^{c\times\hat{b}}$ are trainable transformation matrices.
Then, we obtain the multi-interest representation matrix $\mathbf{M} \in \mathbb{R}^{c\times d}$ as 
\begin{equation}
    \mathbf{M} = \mathbf{A}\mathbf{E}.
\end{equation}
Thus, for every user, we adopt $c$ representation vectors to capture her/his diverse interests.

\subsection{Dependencies Among Captured Interests}
\label{section:DependenciesAmongCapturedInterests}
Prior multi-interest networks capture $c$ representations for user's multiple interests through the multi-interest extractor.
When they have been deployed to serve online traffic, the corresponding items of $c$ captured correlated interests are usually retrieved together to get exposure and collected into training data, known as the sample selection bias problem \cite{ma2018entire,yuan2019improving,chen2020esam}.
Model may incorrectly concentrate on the dependencies among $c$ interests instead of the true causation between \emph{relevant interests} and target items.
Unfortunately, in real scenarios, due to the rapid changing of recommender systems and the temporal evolution of user interest, the test distribution shifts from the training distribution.
To achieve stable performance under such distribution shift, we have to make model focus on the true causation.

To alleviate above problem, we first need to measure the degree of independence between any pair of captured interests $\mathbf{M}_{i,:}$ and $\mathbf{M}_{j,:}$ in the high-dimensional representation space, which is infeasible to resort to histogram-based measures.
In this paper, we introduce the widely used HSIC \cite{gretton2005measuring,gretton2007kernel} to be such independence testing statistics, which is the Hilbert-Schmidt norm of the cross-covariance operator between the distributions in Reproducing Kernel Hilbert Space (RKHS).
For two random variables $U$ and $V$, the formulation of HSIC is:
\begin{equation}
    \begin{aligned}
    HSIC(U, V)& = \mathbb{E}_{uu'vv'}[k_u(u, u')k_v(v, v')] \\
    &+ \mathbb{E}_{uu'}[k_u(u, u')]\mathbb{E}_{vv'}[k_v(v,v')] \\
    &- 2\mathbb{E}_{uv}[\mathbb{E}_{u'}(k_u(u, u'))\mathbb{E}_{v'}[k_v(v,v')]],
    \end{aligned}
\end{equation}
where $\mathbb{E}_{uu'vv'}$ denotes the expectation over independent pairs $(u, v)$ and $(u', v')$ drawn from $P(U, V)$, $k_u$ and $k_v$ are kernel functions.
We use the Radial Basis Function (RBF) kernel which is formulated as:
\begin{equation}
    k(u,v) = exp(-\frac{||u-v||^2_2}{\sigma^2}).
\end{equation}
Given $m$ samples drawn from $P(U, V)$, the Empirical HSIC \cite{gretton2005measuring} is defined as
\begin{equation}
    HSIC(U,V) = (m-1)^{-2}tr(\mathbf{K}_U\mathbf{P}\mathbf{K}_V\mathbf{P}),
\end{equation}
where $\mathbf{K}_U \in \mathbb{R}^{m\times m}$ and $\mathbf{K}_V \in \mathbb{R}^{m\times m}$ have entries $\mathbf{K}_{U_{ij}} = k(U_i,U_j)$ and $\mathbf{K}_{V_{ij}} = k(V_i,V_j)$, and $\mathbf{P} = \mathbf{I} - \frac{1}{m}\mathbf{1}\mathbf{1} \in \mathbb{R}^{m\times m}$ is the centering matrix.
To be noted, $HSIC(U, V) = 0$ if and only if $U \bot V$.

As shown in Figure \ref{fig:baseline_hsic_recall}, taking ComiRec-SA \cite{cen2020controllable} as an example, in training phase, we record the averaged HSIC value among captured interests' representations and the recall on validation set.
Due to the random initialization of model parameters, the value of HSIC and recall are nearly zero at first step, and they rapidly rise along with the increase of training steps.
Around 10000 steps, HSIC slightly declines, but recall keeps rising.
After 20000 steps, the value of HSIC begins to increase slowly while the value of recall stops rising and even begins to decline.
To some extent, for multi-interest networks, this example reveals the trade-offs between the dependencies among interests and model performance on validation set.

As the HSIC is differentiable for back-propagation, to eliminate the influence of dependencies among captured interests, some prior work \cite{bahng2020learning} directly adds it to the original task loss and solve the weighted overall loss by alternative updates.
However, as shown in Figure \ref{fig:baseline_hsic_recall}, the value of HSIC and recall rise together in the early stage of training.
Directly alternating minimizing HSIC and original task loss may increase the instability of training and slow down the convergence speed.
Thus, we need to find more "soft" way to eliminate the dependencies among captured interests and make model concentrate more on underlying true causation between interests and targets.

\subsection{Training \& Serving}
Inspired by sample reweighting techniques \cite{kuang2020stable,zhang2021deep}, we propose a interest decorrelation regularizer which aims to estimate a weight for each sample such that interests are decorrelated on the weighted training data.
Specifically, let $\mathbf{w} \in \mathbb{R}^{n}_{+}$ be the set of sample weight for training data where $n$ is the number of training samples.
We use $\mathbf{w}^{(q)}$ to denote sample weights after the calculation in training epoch $q$, and initialize samples weights as ones, i.e., $\mathbf{w} \in \mathbf{1}^{n}$.

Given the $h$-th training sample with user $u$, let $\mathbf{V}_{\mathcal{S}^{u}_{t+1}}$ denotes the embedding of the target item and $\mathbf{M} \in \mathbb{R}^{k\times d}$ be the corresponding captured interests' representations for the input sequence $\mathcal{S}^{u}_{t}$.
Specifically, we adopt the same interest selection in \cite{cen2020controllable} to choose a interest representation from captured interests to represent user embedding $\mathbf{M}_u$, which can be formulated as
\begin{equation}
    \mathbf{M}_u = \mathbf{M}[\mathop{argmax}(\mathbf{M}\mathbf{V}_{\mathcal{S}^{u}_{t+1}}^{\top}),:].
\end{equation}
Then, the original objective function for the $h$-th training sample can be formulated as
\begin{equation}
    \mathcal{L}_h = -log(\frac{\mathop{exp}(\mathbf{M}_u\mathbf{V}_{\mathcal{S}^{u}_{t+1}}^{\top})}{\sum_{i\in \mathcal{I}}\mathop{exp}(\mathbf{M}_u\mathbf{V}_i^\top)}),
\end{equation}
which can be implemented by the sampled softmax technique \cite{covington2016deep,jean2014using} considering computational efficiency.
At training epoch $q$, applying sample weights for weighted training, the objective function $\hat{\mathcal{L}}_h$ of the proposed DESMIL can be formulated as
\begin{equation}
    \label{weighted_objective_function}
    \hat{\mathcal{L}}_h^{(q)} = \mathbf{w}_h^{(q-1)}\mathcal{L}_h.
\end{equation}

Then, we need to calculate sample weight $\mathbf{w}_h^{(q)}$ for decorrelating interests in above model.
The previous sample weight $\mathbf{w}_h^{(q-1)}$ of the $h$-th training sample is exploited to reweight $\mathbf{M}$ as
\begin{equation}
    \hat{\mathbf{M}} = \mathbf{w}_h^{(q-1)}\mathbf{M}.
\end{equation}
Then, we can find the new optimal sample weight $\mathbf{w}_h^{(q)}$ that minimizes dependency among captured interests of $h$-th training sample as
\begin{equation}
    \label{w_h_argmin}
    \mathbf{w}_h^{(q)} = \mathop{argmin}\limits_{\mathbf{w}_h}\sum_{i}\sum_{j}\lambda HSIC(\hat{\mathbf{M}}_{i,:}, \hat{\mathbf{M}}_{j,:}),
\end{equation}
where $\lambda$ is the decorrelation importance which controls the convergence rate of $\mathbf{w}$.
To be noted, Eq \eqref{w_h_argmin} can be easily implemented via tensorflow \footnote{\url{https://www.tensorflow.org/}\label{fn_tensorflow}} to support batch training efficiently.

\begin{algorithm}
    \caption{Training process of DESMIL}
    \label{alg_training}
    \begin{algorithmic}[1]
        \REQUIRE Training dataset $\mathcal{D}_{tr}$,  maximum training epoch $\mathop{Epoch}$ and batch size $\mathop{B}$.
        \ENSURE Model parameters $\theta$.
        \STATE Initialize the iteration variable $q\gets 0$.
        \STATE Initialize the best iteration variable $q_{best}\gets 0$.
        \STATE Initialize sample weights $\mathbf{w}^{(0)}\gets \mathbf{1}^{n}$.
        \STATE Initialize model parameters $\theta^{(0)}$ via glorot uniform initializer \cite{glorot2010understanding}.
        \REPEAT
        \STATE Extract next batch from $\mathcal{D}_{tr}$.
        \STATE $q\gets q+1$.
        \STATE Keeping $\mathbf{w}^{(q-1)}$ fixed and optimizing $\sum\limits_h^{\mathop{B}} \hat{\mathcal{L}}_h^{(q)}$ via updating $\theta^{(q)}$, where $\hat{\mathcal{L}}_h^{(q)}$ is defined in Eq. \eqref{weighted_objective_function}.
        \STATE  Keeping $\theta^{(q)}$ fixed and updating the corresponding batch sample weights $\mathbf{w}^{(q)}_{\mathop{B}}$ via Eq. \eqref{w_h_argmin} on samples in the batch.
        \STATE Update $q_{best}\gets q$, if better validation result achieved.
        \UNTIL{early stopped or maximum training epoch is reached.}
        \RETURN $\theta^{(q_{best})}$.
    \end{algorithmic}
\end{algorithm}

To be noted, we alternatively update the sample weights via Eq.\eqref{w_h_argmin} and the model parameters via optimizing $\hat{\mathcal{L}}_h^{(q)}$ with respect to model parameters $\theta$.
The detailed procedure of our model is shown in Algorithm \ref{alg_training}.

\begin{table*}[t]
\centering
\caption{Results on Public and industrial Datasets. Best performances are indicated by bold fonts and the strongest baselines are underlined. The improvement (Improv.) indicates the relative increase of our model over baselines on metrics. }
  \label{tab:public and industrial result}
\begin{tabular}{ccccccccccc}
\toprule
Datasets & Metric & POP & GRU4Rec & Y-DNN & SASRec & MIND & ComiRec & CauseRec & DESMIL & Improv.  \\
\midrule
\multirow{6}{*}{Book} 
 & Recall@20 & 1.37&3.47&4.40&4.76&5.10&\underline{5.92}&5.75&\textbf{7.52} & 27.03\%  \\
 & Recall@50 &2.40&6.50&7.31&7.78&7.64&\underline{9.35}&\underline{9.36}&\textbf{11.06} & 18.16\%   \\
 & NDCG@20 &2.26&3.55&4.59&4.84&\underline{5.09}&4.17&4.66&\textbf{5.46}  & 7.27\%  \\
 & NDCG@50 &3.94&4.42&5.54&5.74&5.97&5.47&\underline{6.28}&\textbf{7.24} & 15.28\%  \\
 & HR@20   &3.02&7.84&9.89&8.82&10.59&11.70&\underline{12.45}&\textbf{14.86} & 19.36\%   \\
 & HR@50   &5.23&12.38&14.94&13.79&15.56&18.04&\underline{20.23}&\textbf{21.53}  & 6.43\%  \\
 \midrule
\multirow{6}{*}{Movies and TV} 
 & Recall@20 &3.59&13.20&12.38&14.43&14.87&\underline{15.46}&15.30&\textbf{15.76}  & 1.94\%  \\
 & Recall@50 &6.62&17.66&17.31&18.27&\underline{19.55}&18.87&19.24&\textbf{20.90} & 6.91\%  \\
 & NDCG@20 &5.30&15.07&12.64&14.49&\underline{15.80}&14.73&15.10&15.31&  \verb|\|  \\
 & NDCG@50 &9.66&16.21&14.11&16.72&\underline{17.23}&16.17&16.83&\textbf{17.36} & 0.75\%   \\
 & HR@20   &6.51&22.67&21.32&23.25&25.34&25.87&\underline{25.94}&\textbf{26.42}&  1.85\%  \\
 & HR@50   &11.73&29.54&29.46&30.43&32.93&33.68&\underline{33.90}&\textbf{34.80} & 2.65\%   \\
 \midrule
\multirow{6}{*}{CDs and Vinyl} 
 & Recall@20 &0.993&4.39&5.24&6.92&7.55&\underline{7.96}&7.77&\textbf{8.75}&  9.92\%  \\
 & Recall@50 &1.89&6.07&7.72&8.52&10.32&\underline{11.23}&11.12&\textbf{12.09} & 7.66\%  \\
 & NDCG@20   &1.58&4.81&5.42&6.44&\underline{7.93}&6.84&7.51&7.79&  \verb|\|  \\
 & NDCG@50   &3.12&5.42&6.36&7.10&\underline{8.88}&8.01&8.57&\textbf{8.86} & \verb|\|  \\
 & HR@20     &2.11&8.47&10.40&12.86&14.28&14.35&\underline{14.49}&\textbf{15.73}& 8.56\%  \\
 & HR@50     &4.08&11.79&15.46&16.29&19.38&20.26&\underline{20.66}&\textbf{21.89} & 5.95\%  \\
 \midrule
\multirow{6}{*}{Industrial Dataset} 
 & Recall@20 &1.01&6.31&6.28&6.81&6.96&\underline{7.23}&7.02&\textbf{8.41}&  16.32\%  \\
 & Recall@50 &1.75&9.88&10.76&11.02&11.29&11.51&\underline{11.76}&\textbf{12.87}& 9.44\%  \\
 & NDCG@20   &1.92&9.15&7.25&8.27&\underline{8.91}&8.58&8.60&\textbf{9.19}&  3.14\%  \\
 & NDCG@50   &3.28&10.60&9.20&9.78&\underline{10.75}&10.35&10.54&\textbf{11.28}&  4.93\%  \\
 & HR@20     &2.55&17.04&16.73&17.05&17.52&17.89&\underline{18.33}&\textbf{20.50}&  11.84\%  \\
 & HR@50     &4.38&24.97&23.94&25.11&26.21&26.56&\underline{26.77}&\textbf{29.45} & 10.01\%  \\
\bottomrule
\end{tabular}
\end{table*}

At serving time, sample weights are not involved and only the backbone model that generates the user representation is needed, producing multiple representation vectors for each user.
Each interest embedding can independently retrieve top-N items based on an approximate nearest neighbor approach \cite{johnson2019billion}.
These items constitute the final set of candidate items for the matching stage of recommender systems.

\section{EXPERIMENTS}
In this section, we perform extensive experiments to evaluate the performance of our model.
Firstly, we introduce datasets, comparison models and metrics used for evaluation. Then, quantitative results and visualization are discussed to empirically analyze the proposed DESMIL.

\begin{figure*}
    \centering
    \includegraphics[width=0.7\linewidth]{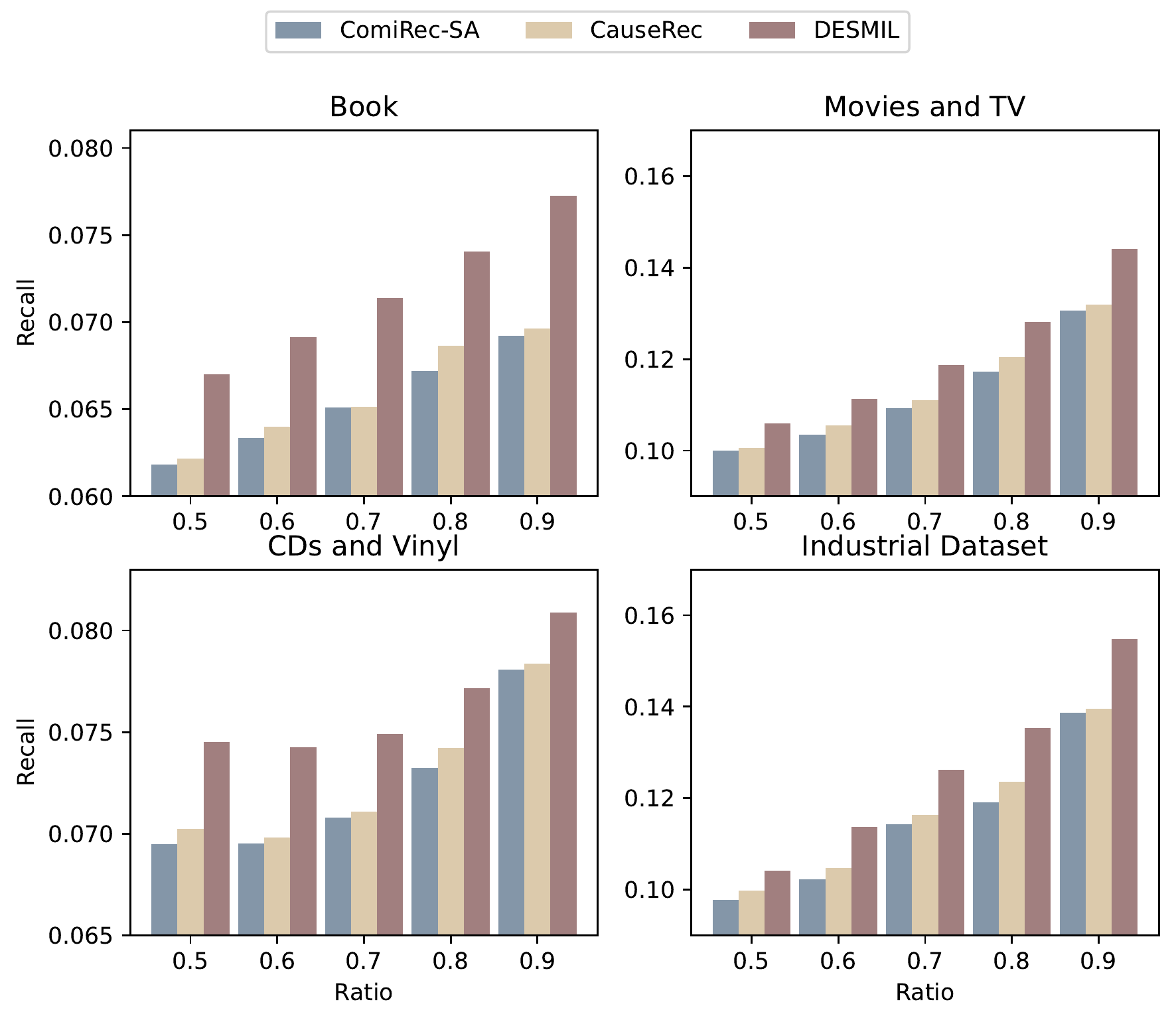}
    \caption{Comparison result of Recall@50 on four datasets with different ratio of simulated covariate shift.}
    \label{fig:ood_recall}
\end{figure*}

\subsection{Experimental Setup}
\label{section:ExperimentalSetup}
\textbf{Datasets}. We evaluate our proposed model on three public datasets and a large-scale commercial datasets as described in the following. 

\begin{itemize}
\item \textbf{Book Dataset}. The book dataset is part of the Amazon product data\footnote{\url{http://jmcauley.ucsd.edu/data/amazon/}} in the "book" category. This dataset is introduced in \cite{mcauley2015image, he2016ups}, which consists of product reviews spanning May 1996 to July 2014 from Amazon.com. There are 603668 users, 367982 items and 8898041 user behaviors in total.

\item \textbf{Movies and TV Dataset}. The Movies and TV dataset is part of the updated version of Amazon Review Data\footnote{\url{https://nijianmo.github.io/amazon/index.html}} \cite{ni2019justifying}. This dataset contains product reviews in the "Movies and TV" category from May 1996 to Oct 2018. There are 304763 users, 89590 items, and 3506470 user behaviors in total.

\item \textbf{CDs and Vinyl Dataset}. The CDs and Vinyl dataset is also part of the updated Amazon Review Data \cite{ni2019justifying}. This dataset contains product reviews in the "CDs and Vinyl" category from May 1996 to Oct 2018. There are 129237 users, 145522 items, and 1682049 user behaviors in total.

\item \textbf{Industrial Dataset}. The Industrial dataset is sampled from real-world mobile application logs, and all records have been anonymized and sanitized. This dataset has 5886272 users, 1195809 items and 103579419 user behaviors in total.
\end{itemize}
For the classic dataset splitting, the Book dataset follows the official repository \footnote{\url{https://github.com/THUDM/ComiRec}} of ComiRec.
For the OOD splitting, for each user $u$, if the length of the sequence of historical behaviors $\mathcal{S}^u$ is less than $10$, $\mathcal{S}^u$ will be partitioned to training set only.\\
\textbf{Competitors}. We compare our proposed model to the following baselines for evaluation.
\begin{itemize}
\item \textbf{POP}: a simplest baseline that ranks items according to their popularity (i.e., the number of interactions).
\item \textbf{GRU4Rec} \cite{hidasi2015session}: an early sequential recommendation model based on recurrent neural network.
\item \textbf{Y-DNN} \cite{covington2016deep}: one of the most successful deep learning models for industrial recommender systems.
\item \textbf{SASRec} \cite{kang2018self}: a state-of-the-art model that uses self-attention network for the sequential recommendation.
\item \textbf{MIND} \cite{li2019multi}: a state-of-the-art multi-interest sequential recommendation model with dynamic routing for modeling user’s diverse interests in the matching stage.
\item \textbf{ComiRec} \cite{cen2020controllable}: a state-of-the-art sequential recommendation model with multi-interest extraction module to generate multiple user interests and aggregation module to obtain top-N items. We use the SA setting of ComiRec which is described as ComiRec-SA in the original paper.
\item \textbf{CauseRec} \cite{zhang2021causerec}: a state-of-the-art sequential recommendation model that performs contrastive user representation learning to model the counterfactual data distribution to confront the sparsity and noise nature of observed
user interactions.
\end{itemize}
Meanwhile, we implement our proposed DESMIL model with TensorFlow 1.14 and Faiss \footnote{\url{https://github.com/facebookresearch/faiss}\label{fn_faiss}} in Python 3.7. Experiments on public datasets are conduct using a single Linux server with 4 Intel(R) Xeon(R) CPU E5-2680 v4@ 2.40GHz, 256G RAM, and 8 NVIDIA GeForce RTX 2080 Ti.\\
\textbf{Parameter Configuration}.
The dimension of item embedding is set to 64.
The batch size for Book dataset and Industrial Dataset is set to 1024, while the batch size for the Movies and TV dataset and the CDs and Vinyl dataset is set to 128, according to the best performances of ComiRec.
The number of negative samples for sampled softmax loss is set to 10.
All models use early stopping based on the Recall@50 on the validation set.
The decorrelation importance and the number of interests are tuned in the range of $\left\{0.01, 0.1, 1.0, 10.0, 100.0\right\}$ and the range of $\left\{2, 4, 6, 8\right\}$, respectively.
We use the Adam optimizer \cite{kingma2014adam} with learning rate lr = 0.001 for optimization.\\
\textbf{Evaluation Metrics}. We use the top $p$ Recall Rate, Normalized Discounted Cumulative Gain (NDCG), and Hit Rate (HR) to evaluate the sequential recommendation model performance. We select $p=20, 50$ in the experiments. The three metrics measure the model performance with different criteria. Recall@$p$ is often considered the most important factor for a model to be used in the industry.
Recall@$p$ is defined as the fraction of relevant items found in the top $p$ recommended items. NDCG@$p$ further considers the normalization of gains and the ranking of correctly recommended items, where items with higher relevance affect the final score more. HR@$p$ is defined as the proportion of top $p$ recommended samples found in the test set. 

\subsection{Results and Analysis}
\label{section:ResultsAndAnalysis}

Firstly, we conduct experiments on dataset splitting following previous work \cite{cen2020controllable}.
Specifically, we split samples into training, validation and test set according to the corresponding users (namely classic splitting).
Considering interaction time of samples in training, validation and test set may overlap, in such kind of data splitting, samples in different sets share relative similar data distribution.
In Table \ref{tab:public and industrial result}, we show the performance comparison on the four datasets.
All the presented results are averaged by five independent results with different seeds.
From the table, we can tell that DESMIL, CauseRec, ComiRec and MIND outperform earlier deep learning models such as SASRec, Y-DNN and GRU4Rec. And POP performs poorly.
Among the three state-of-the-art models, MIND performs relatively better on NDCG, ComiRec performs relatively better on recall rates, CauseRec performs relatively better on hit rates.
To be noted, as we follow the correct calculation of NDCG which is corrected by the authors in the official repository of ComsiRec \cite{cen2020controllable} to evaluate each model, the NDCG results of ComsiRec and CauseRec are lower than what are reported in the original papers.
Moreover, according to results in Table \ref{tab:public and industrial result}, our model relatively outperforms previous state-of-the-art models by 1.94\% to 27.03\% in Recall and by 1.85\% to 19.36\% in Hit Rate.
These improvements are significant.
When evaluated by the metric of NDCG, our proposed model outperforms other models on the Book dataset and the Industrial dataset by large margins from 3.14\% to 15.28\%.
But on the Movies and the TV dataset and the CDs and Vinyl dataset, DESMIL achieves very slightly lower but acceptable NDCG results, compared with MIND.
In real applications, Recall is often considered as the most important metric as it can best reflect the performance of real-world recommender systems facing enormous candidate set of items and almost equally important but limited exposure positions.
Comprehensively considering the importance of Recall and the significant improvements measured by Recall and Hit Rate, DESMIL still greatly outperforms other compared models, and achieves promising performances on all the four datasets.

\begin{figure}
    \centering
    \includegraphics[width=0.95\linewidth]{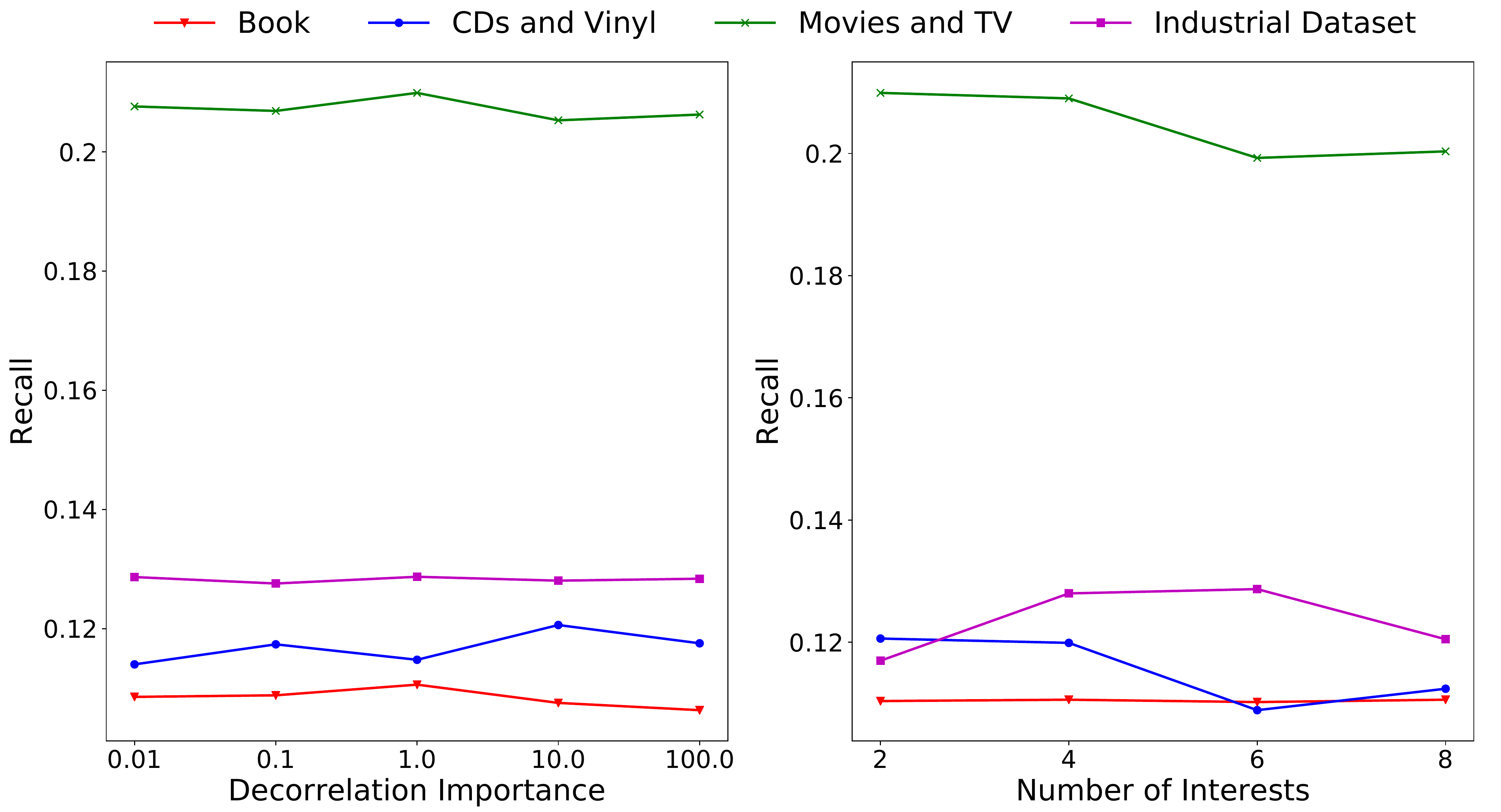}
    \caption{Hyperparameter study on decorrelation importance coefficient and number of interests measured by Recall@50.}
    \label{fig:weight_recall}
\end{figure}
\begin{figure}
    \centering
    \includegraphics[width=0.95\linewidth]{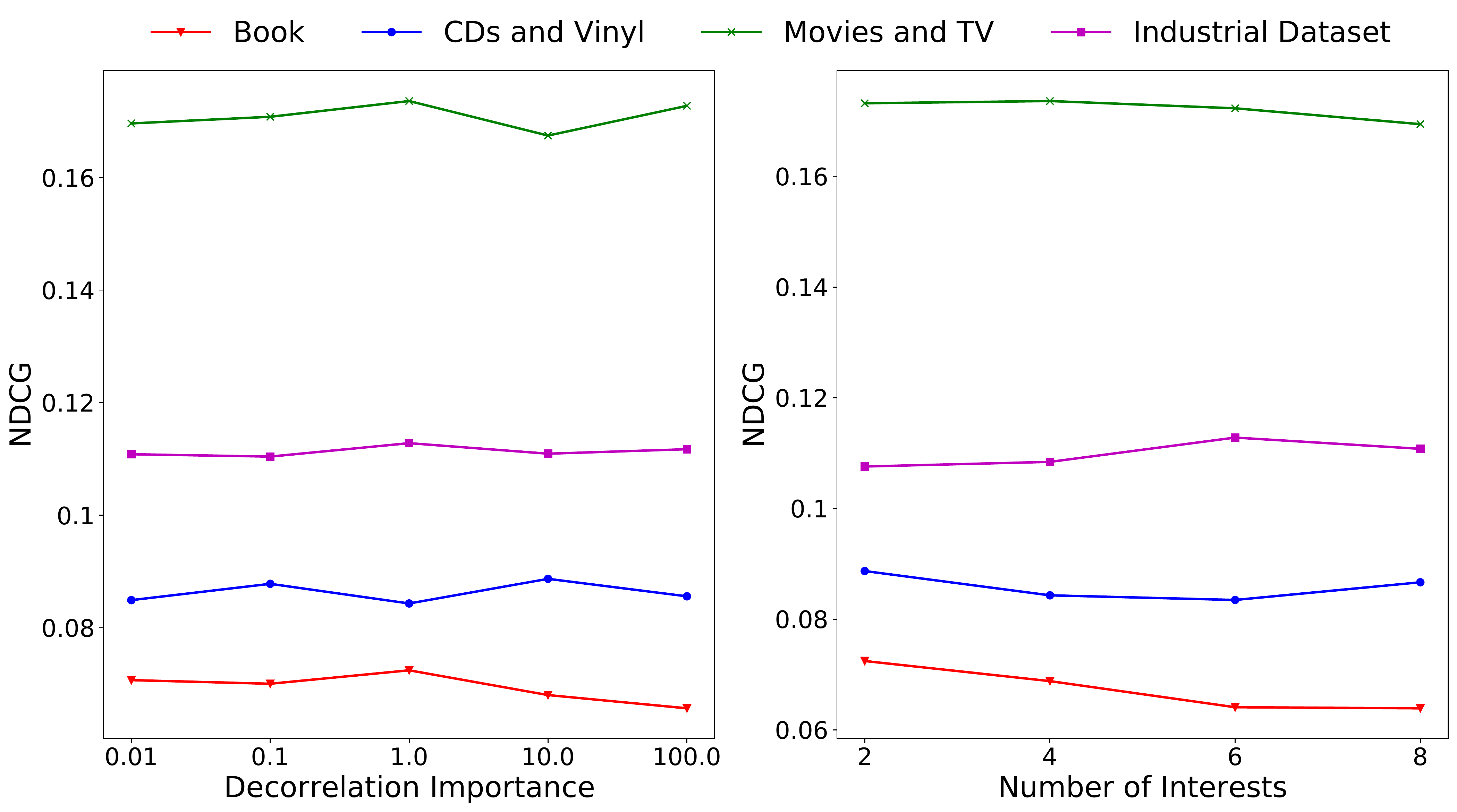}
    \caption{Hyperparameter study on decorrelation importance coefficient and number of interests measured by NDCG@50.}
    \label{fig:weight_ndcg}
\end{figure}
\begin{figure}
    \centering
    \includegraphics[width=0.95\linewidth]{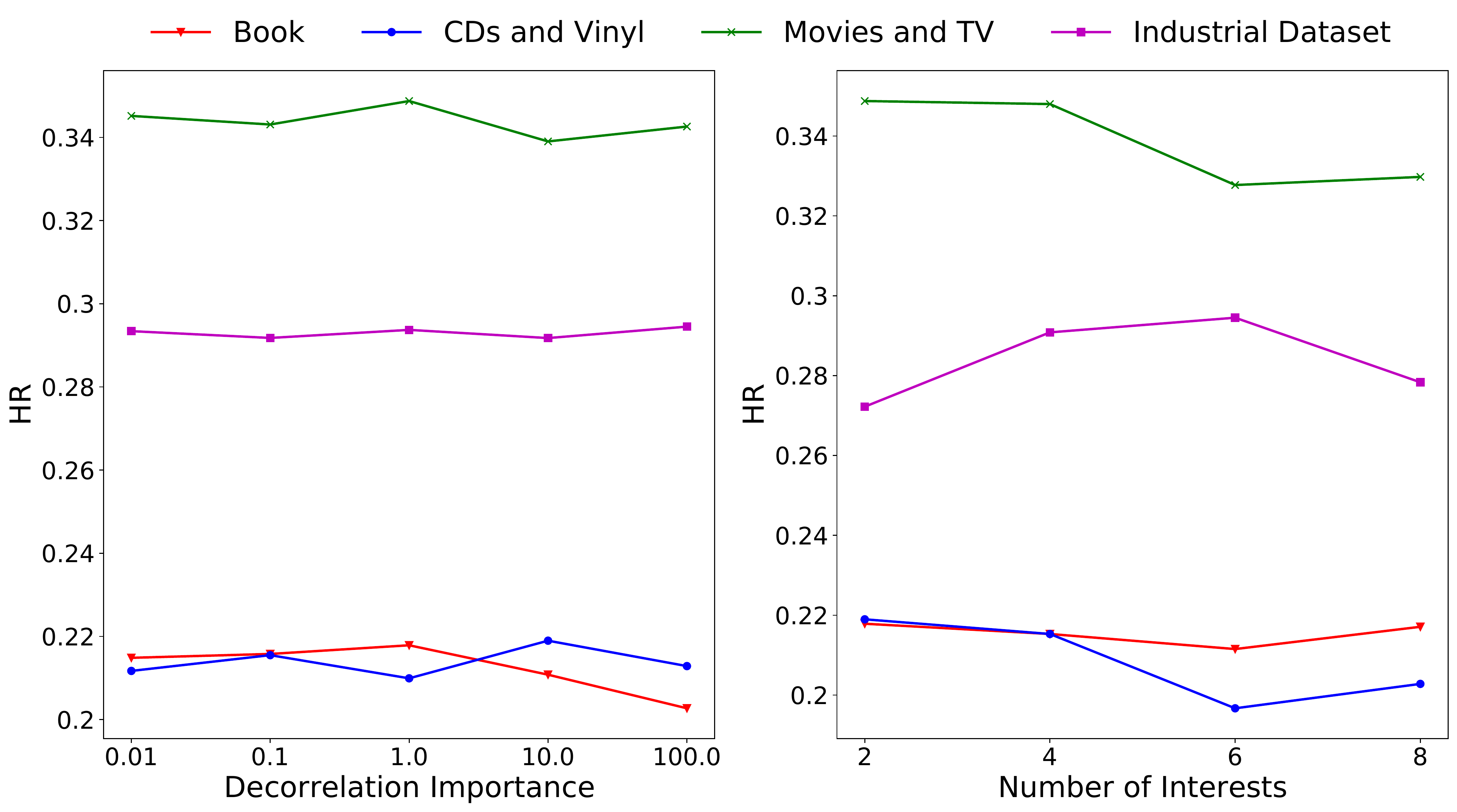}
    \caption{Hyperparameter study on decorrelation importance coefficient and number of interests measured by HR@50.}
    \label{fig:weight_hr}
\end{figure}

Secondly, we conduct performance comparison on OOD data, for investigating whether recommendation models can provide stable predictions.
Considering data distribution and dependencies among captured interests may change in different time periods, we conduct the following data splitting (namely OOD splitting).
We use the first 50\% and the following 10\% samples in sequences as training set and validation set respectively. 
After training on above data is completed, we adopt the first $z$ of samples and the left $1-z$ of samples in sequences as input historical behaviors and the test samples, respectively.
To simulate different ratio of covariate shift \cite{xu2021stable,shen2021towards}, $z$ takes value in the range of $\left\{0.5, 0.6, 0.7, 0.8, 0.9\right\}$.
In Figure \ref{fig:ood_recall}, we illustrate performance comparison on such OOD datasets with different ratio of simulated covariate shift.
For simplicity, we only show results evaluated by Recall@50.
To be noted, compared with \cite{cen2020controllable}, such data splitting is more applicable and reasonable considering real-world recommender systems, for we usually train models based data collected before a time point, and use the models online for predicting samples after the time point.
We can clearly observe from the figure that, DESMIL stably outperforms the state-of-the-art models ComsiRec and CauseRec by large margins.
These results further show the effectiveness and stability of DESMIL under distribution shifts.

Thirdly, we need to investigate the impact of some key hyperparameters.
We conduct experiments on datasets under the classic splitting, which stays the same as in Table \ref{tab:public and industrial result}.
We illustrate hyperparamter study measured by Recall@50, NDCG@50 and HR@50 in Figure \ref{fig:weight_recall}, \ref{fig:weight_ndcg} and \ref{fig:weight_hr} respectively.
Similar phenomenons are shared across figures measured by different evaluation metrics.
As shown in the figures, the selection of the decorrelation importance coefficient $\lambda$ to control the convergence rate of training sample weight does not affect the model performances very much. So we can simply set $\lambda=1.0$ for most datasets.
Meanwhile, the optimal number $c$ of interests may vary among different datasets. The optimal number of interests for the Book dataset, the CDs and Vinyl dataset and the Movies and TV dataset is $c=2$, and for Industrial dataset is $c=6$.
Overall speaking, the performance of DESMIL with varying hyperparameters is relatively stable, especially in some ranges of the hyperparameters. This does not leave us too much burden for hyperparameter tuning in practice.

\subsection{Visualization}
\begin{figure}
    \centering
    \includegraphics[width=0.95\linewidth]{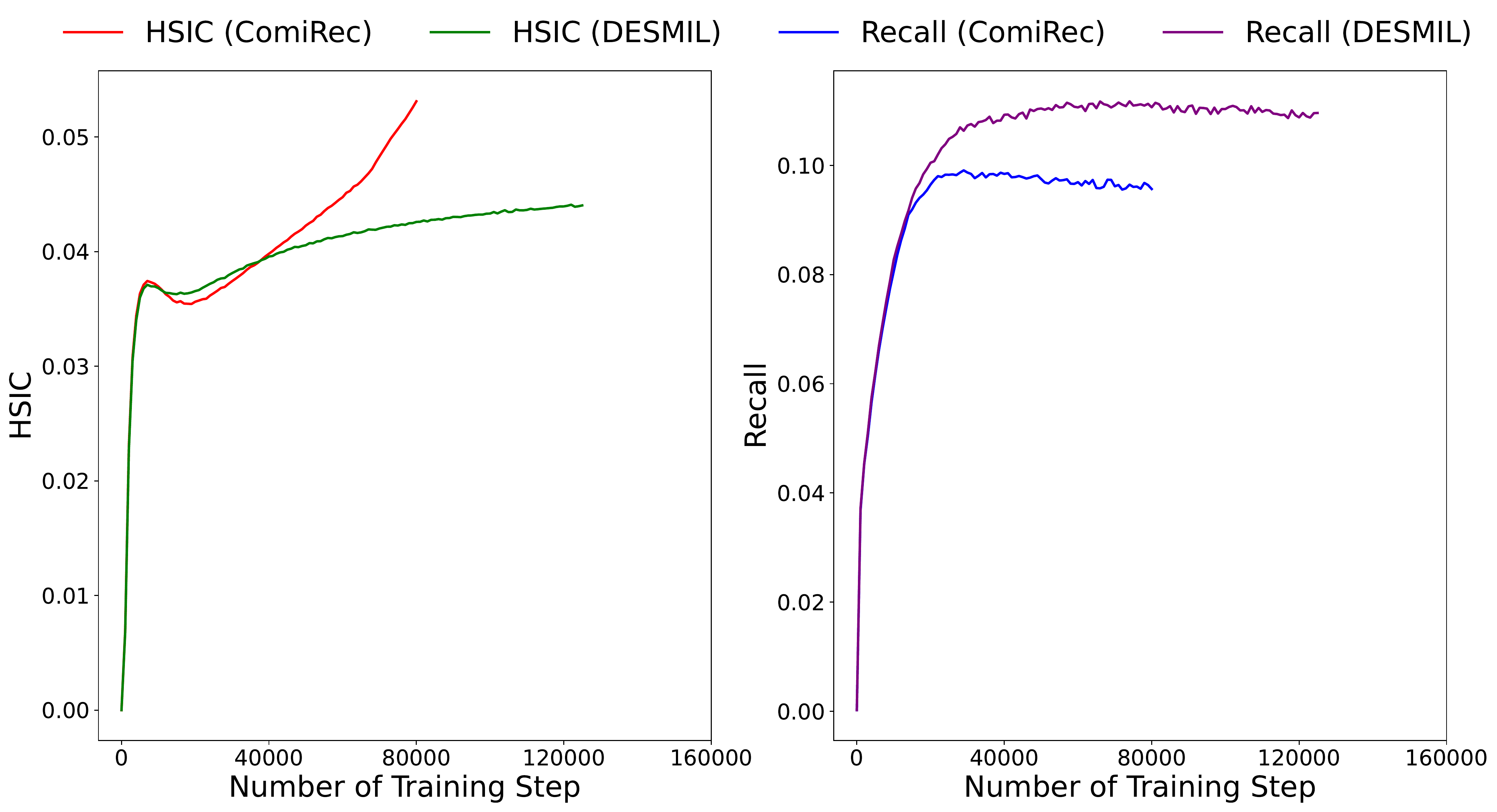}
    \caption{The curves of HSIC on the training set and Recall@50 on the validation set when training ComiRec and DESMIL on the Book dataset. With the use of early stopping, the training of them terminates at different step, which results in the different length of curves.}
    \label{fig:insight_book}
\end{figure}

\begin{figure}
    \centering
    \includegraphics[width=0.95\linewidth]{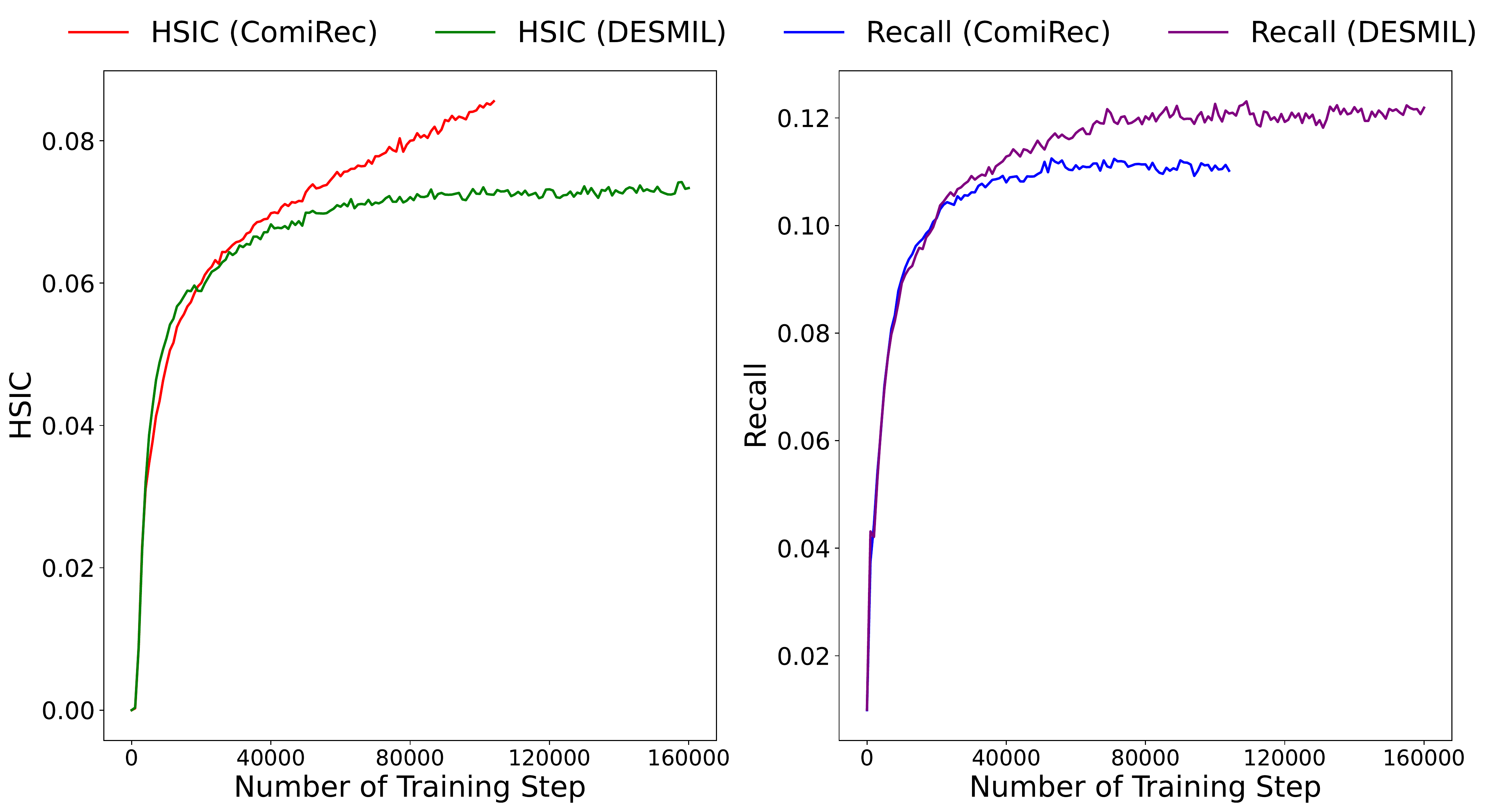}
    \caption{The curves of HSIC on the training set and Recall@50 on the validation set when training ComiRec and DESMIL on the CDs and Vinyl dataset.}
    \label{fig:cds_hsic_recall}
\end{figure}

\begin{figure}
    \centering
    \includegraphics[width=.95\linewidth]{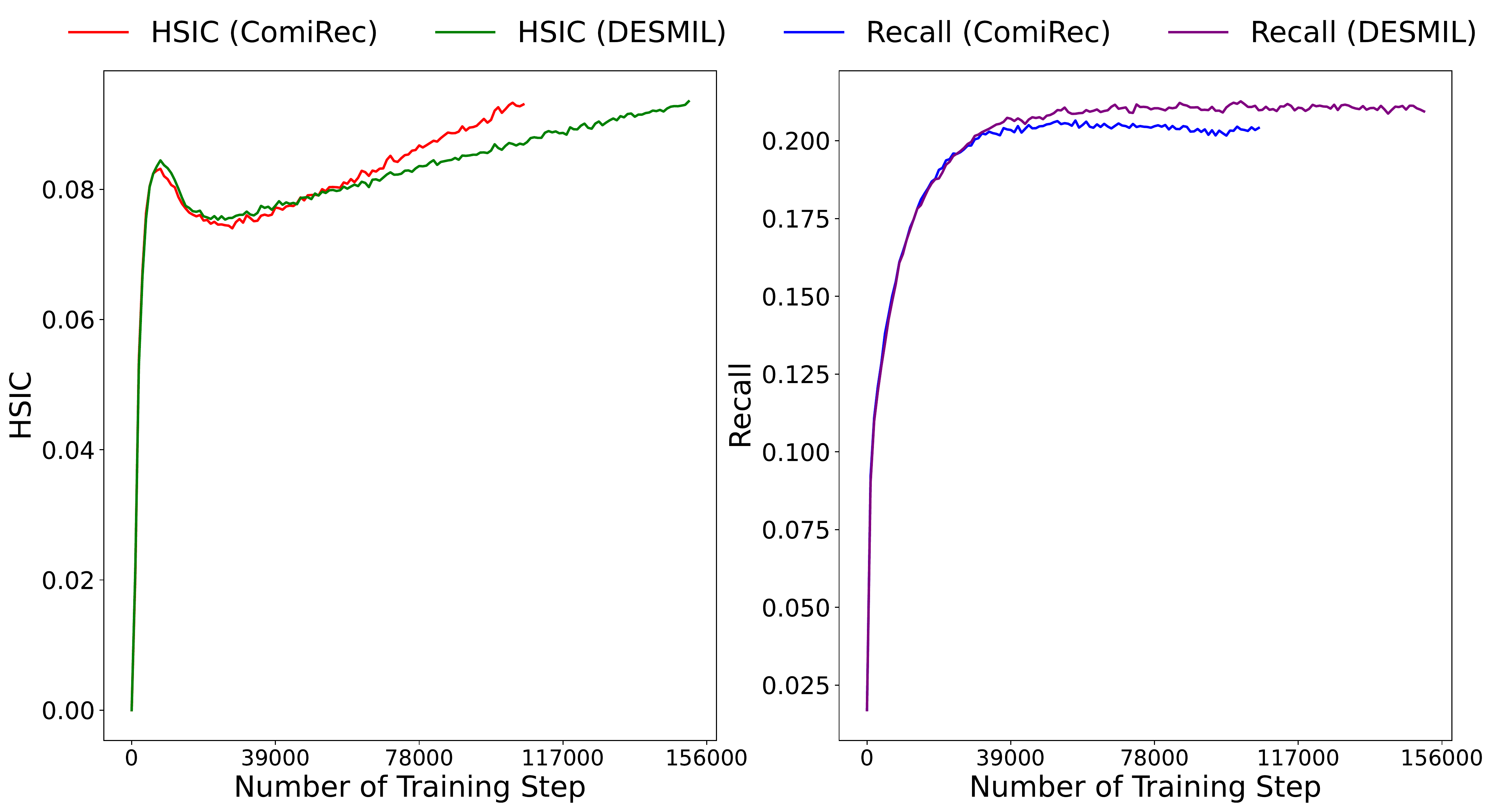}
    \caption{The curves of HSIC on the training set and Recall@50 on the validation set when training ComiRec and DESMIL on the Movies and TV dataset.}
    \label{fig:mat_hsic_recall}
\end{figure}

In Figure \ref{fig:insight_book}, we visualize the change of HSIC on the training set and Recall on validation set when training DESMIL and ComiRec on the Book dataset. Both DESMIL and ComiRec use early stopping and the training of them terminates at different step, which results in the different length of curves shown in Figure \ref{fig:insight_book}. It should be noted that DESMIL performs optimization of HSIC by sample weighting in the training phase, while the calculation of HSIC, which is shown in Figure \ref{fig:insight_book}, is not weighted. Different from DESMIL, ComiRec does not control the optimization of HSIC (i.e., dependencies among interests) at all. At the first 10000 steps, the HSIC and Recall of both models increase. Then, the HSIC of ComiRec continues to rise while the HSIC of DESMIL is well under control, which makes it possible for DESMIL to update more steps and obtain better performance.
Similar as the Book dataset, we further illustrate the curves of HSIC on the training set and Recall on validation set when training DESMIL and ComiRec on the CDs and Vinyl dataset and the Movies and TV dataset.
As shown in Figure \ref{fig:cds_hsic_recall}, for the CDs and Vinyl dataset, after the first $40000$ steps, the HSIC of DESMIL is well under control while the HSIC of ComiRec continues to rise.
As shown in Figure \ref{fig:mat_hsic_recall}, for the Movies and TV dataset, the HSIC of DESMIL continues to rise while more slower than the HSIC of ComiRec.
From these curves, we can observe a clear correlation between the controlled HSIC and the space for optimization of Recall.

\begin{figure}
    \centering
    \includegraphics[width=0.8\linewidth]{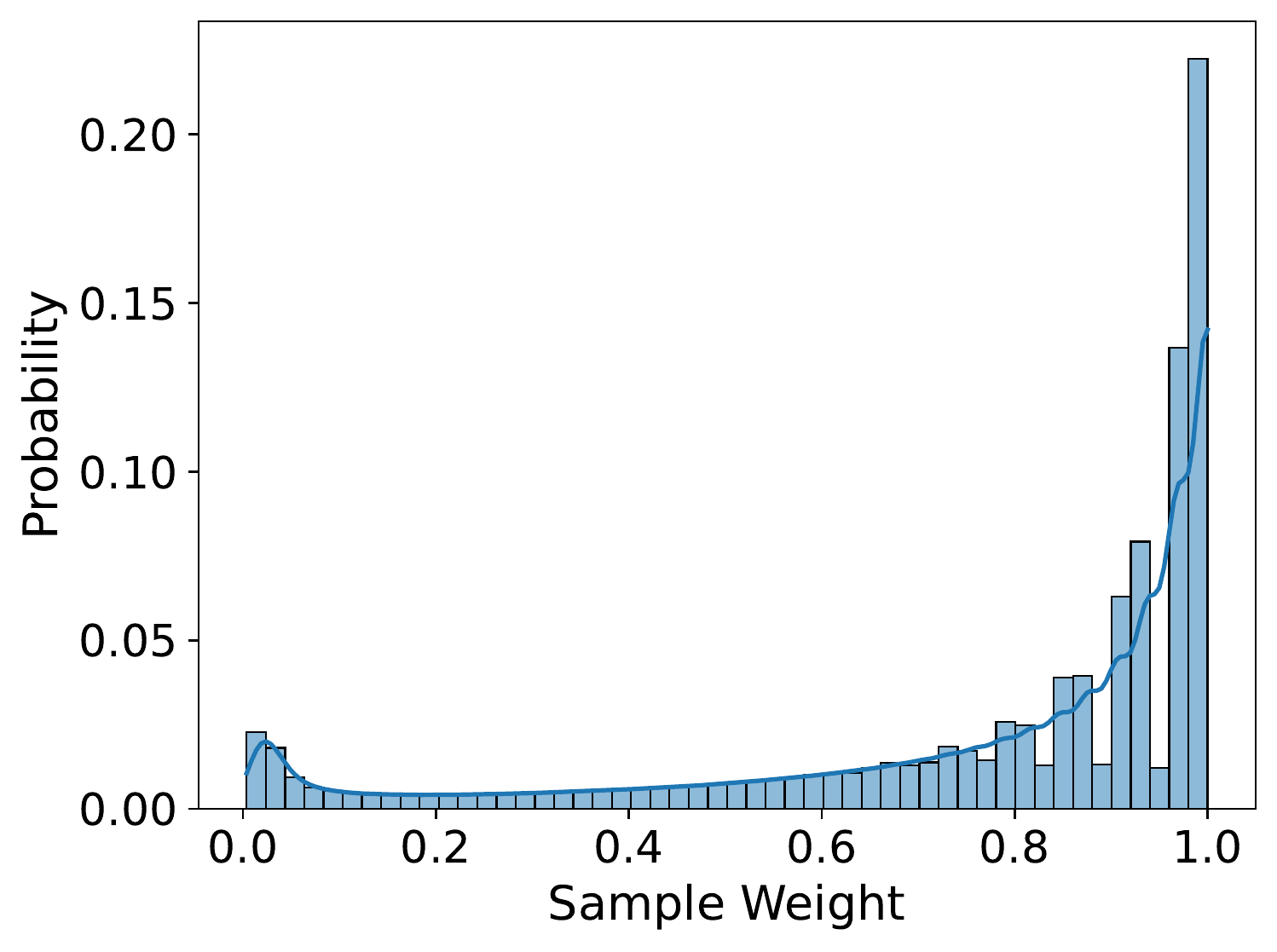}
    \caption{The histogram of sample weight on the Book dataset.}
    \label{fig:insight_histogram}
\end{figure}

In Figure \ref{fig:insight_histogram}, we visualize the probability distribution of sample weights for the Book dataset in a histogram. This shows that the sample weights of user interests are mostly around the values from 0.8 to 1.0, with some located near the value of 0.0. The value near 1.0 means no change of sample weight while the value near 0.0 means a sharp change of sample weight in the loss function. This shows that most data in the Book dataset does not require specific decorrelation techniques while some data points are indeed marginalized. These marginalized data points may decrease the influence of some popular but unfitted user interests and interacted items on the overall model training, and allowing the model to focus on data points that actually possesses causal effect on the prediction even when the environment changes.


\section{Conclusion}
In this paper, for multi-interest networks, we introduce HSIC as independence testing statistics to measure the degree of independence among captured interests.
We trace the HSIC and model performance in training phase, and observe that the continuous increase of HSIC may affect model performance in the middle and late stage of training.
Thus, we point out that eliminating the influence of dependencies among captured interests is a promising way to alleviate OOD generalization problem in recommender systems. 
Based on this, we propose a novel deep stable multi-interest learning for sequential recommendation.
The interest decorrelation regularizer in DESMIL tries to eliminate the influence of subtle dependencies between captured interests via learning weights for training samples, which is a soft way to make model concentrate more on underlying true causation.
Extensive experiments have been conducted to demonstrate that DESMIL achieves superior performance on public benchmarks, large-scale industrial dataset and the synthetic dataset which simulates the OOD data.
Besides, the comprehensive model analysis uncovers the reason why DEMSIL works to a certain extent.

\balance
\bibliographystyle{ACM-Reference-Format}
\bibliography{main}


\begin{thebibliography}{55}


\ifx \showCODEN    \undefined \def \showCODEN     #1{\unskip}     \fi
\ifx \showDOI      \undefined \def \showDOI       #1{#1}\fi
\ifx \showISBNx    \undefined \def \showISBNx     #1{\unskip}     \fi
\ifx \showISBNxiii \undefined \def \showISBNxiii  #1{\unskip}     \fi
\ifx \showISSN     \undefined \def \showISSN      #1{\unskip}     \fi
\ifx \showLCCN     \undefined \def \showLCCN      #1{\unskip}     \fi
\ifx \shownote     \undefined \def \shownote      #1{#1}          \fi
\ifx \showarticletitle \undefined \def \showarticletitle #1{#1}   \fi
\ifx \showURL      \undefined \def \showURL       {\relax}        \fi
\providecommand\bibfield[2]{#2}
\providecommand\bibinfo[2]{#2}
\providecommand\natexlab[1]{#1}
\providecommand\showeprint[2][]{arXiv:#2}

\bibitem[\protect\citeauthoryear{Bahng, Chun, Yun, Choo, and Oh}{Bahng
  et~al\mbox{.}}{2020}]%
        {bahng2020learning}
\bibfield{author}{\bibinfo{person}{Hyojin Bahng}, \bibinfo{person}{Sanghyuk
  Chun}, \bibinfo{person}{Sangdoo Yun}, \bibinfo{person}{Jaegul Choo}, {and}
  \bibinfo{person}{Seong~Joon Oh}.} \bibinfo{year}{2020}\natexlab{}.
\newblock \showarticletitle{Learning de-biased representations with biased
  representations}. In \bibinfo{booktitle}{\emph{ICML}}.
  \bibinfo{pages}{528--539}.
\newblock


\bibitem[\protect\citeauthoryear{Cen, Zhang, Zou, Zhou, Yang, and Tang}{Cen
  et~al\mbox{.}}{2020}]%
        {cen2020controllable}
\bibfield{author}{\bibinfo{person}{Yukuo Cen}, \bibinfo{person}{Jianwei Zhang},
  \bibinfo{person}{Xu Zou}, \bibinfo{person}{Chang Zhou},
  \bibinfo{person}{Hongxia Yang}, {and} \bibinfo{person}{Jie Tang}.}
  \bibinfo{year}{2020}\natexlab{}.
\newblock \showarticletitle{Controllable multi-interest framework for
  recommendation}. In \bibinfo{booktitle}{\emph{KDD}}.
  \bibinfo{pages}{2942--2951}.
\newblock


\bibitem[\protect\citeauthoryear{Chen, Ren, Cai, Sun, and de~Rijke}{Chen
  et~al\mbox{.}}{2020a}]%
        {chen2020improving}
\bibfield{author}{\bibinfo{person}{Wanyu Chen}, \bibinfo{person}{Pengjie Ren},
  \bibinfo{person}{Fei Cai}, \bibinfo{person}{Fei Sun}, {and}
  \bibinfo{person}{Maarten de Rijke}.} \bibinfo{year}{2020}\natexlab{a}.
\newblock \showarticletitle{Improving end-to-end sequential recommendations
  with intent-aware diversification}. In \bibinfo{booktitle}{\emph{CIKM}}.
  \bibinfo{pages}{175--184}.
\newblock


\bibitem[\protect\citeauthoryear{Chen, Xiao, Li, Ye, Sun, and Deng}{Chen
  et~al\mbox{.}}{2020b}]%
        {chen2020esam}
\bibfield{author}{\bibinfo{person}{Zhihong Chen}, \bibinfo{person}{Rong Xiao},
  \bibinfo{person}{Chenliang Li}, \bibinfo{person}{Gangfeng Ye},
  \bibinfo{person}{Haochuan Sun}, {and} \bibinfo{person}{Hongbo Deng}.}
  \bibinfo{year}{2020}\natexlab{b}.
\newblock \showarticletitle{Esam: Discriminative domain adaptation with
  non-displayed items to improve long-tail performance}. In
  \bibinfo{booktitle}{\emph{SIGIR}}. \bibinfo{pages}{579--588}.
\newblock


\bibitem[\protect\citeauthoryear{Covington, Adams, and Sargin}{Covington
  et~al\mbox{.}}{2016}]%
        {covington2016deep}
\bibfield{author}{\bibinfo{person}{Paul Covington}, \bibinfo{person}{Jay
  Adams}, {and} \bibinfo{person}{Emre Sargin}.}
  \bibinfo{year}{2016}\natexlab{}.
\newblock \showarticletitle{Deep neural networks for youtube recommendations}.
  In \bibinfo{booktitle}{\emph{RecSys}}. \bibinfo{pages}{191--198}.
\newblock


\bibitem[\protect\citeauthoryear{Fan, Wang, Shi, Cui, and Wang}{Fan
  et~al\mbox{.}}{2021}]%
        {fan2021generalizing}
\bibfield{author}{\bibinfo{person}{Shaohua Fan}, \bibinfo{person}{Xiao Wang},
  \bibinfo{person}{Chuan Shi}, \bibinfo{person}{Peng Cui}, {and}
  \bibinfo{person}{Bai Wang}.} \bibinfo{year}{2021}\natexlab{}.
\newblock \showarticletitle{Generalizing Graph Neural Networks on
  Out-Of-Distribution Graphs}.
\newblock \bibinfo{journal}{\emph{arXiv preprint arXiv:2111.10657}}
  (\bibinfo{year}{2021}).
\newblock


\bibitem[\protect\citeauthoryear{Fang, Zhang, Shu, and Guo}{Fang
  et~al\mbox{.}}{2020}]%
        {fang2020deep}
\bibfield{author}{\bibinfo{person}{Hui Fang}, \bibinfo{person}{Danning Zhang},
  \bibinfo{person}{Yiheng Shu}, {and} \bibinfo{person}{Guibing Guo}.}
  \bibinfo{year}{2020}\natexlab{}.
\newblock \showarticletitle{Deep learning for sequential recommendation:
  Algorithms, influential factors, and evaluations}.
\newblock \bibinfo{journal}{\emph{ACM Transactions on Information Systems
  (TOIS)}} \bibinfo{volume}{39}, \bibinfo{number}{1} (\bibinfo{year}{2020}),
  \bibinfo{pages}{1--42}.
\newblock


\bibitem[\protect\citeauthoryear{Glorot and Bengio}{Glorot and Bengio}{2010}]%
        {glorot2010understanding}
\bibfield{author}{\bibinfo{person}{Xavier Glorot} {and} \bibinfo{person}{Yoshua
  Bengio}.} \bibinfo{year}{2010}\natexlab{}.
\newblock \showarticletitle{Understanding the difficulty of training deep
  feedforward neural networks}. In \bibinfo{booktitle}{\emph{AISTATS}}.
  \bibinfo{pages}{249--256}.
\newblock


\bibitem[\protect\citeauthoryear{Gretton, Bousquet, Smola, and
  Sch{\"o}lkopf}{Gretton et~al\mbox{.}}{2005}]%
        {gretton2005measuring}
\bibfield{author}{\bibinfo{person}{Arthur Gretton}, \bibinfo{person}{Olivier
  Bousquet}, \bibinfo{person}{Alex Smola}, {and} \bibinfo{person}{Bernhard
  Sch{\"o}lkopf}.} \bibinfo{year}{2005}\natexlab{}.
\newblock \showarticletitle{Measuring statistical dependence with
  Hilbert-Schmidt norms}. In \bibinfo{booktitle}{\emph{COLT}}.
  \bibinfo{pages}{63--77}.
\newblock


\bibitem[\protect\citeauthoryear{Gretton, Fukumizu, Teo, Song, Sch{\"o}lkopf,
  Smola, et~al\mbox{.}}{Gretton et~al\mbox{.}}{2007}]%
        {gretton2007kernel}
\bibfield{author}{\bibinfo{person}{Arthur Gretton}, \bibinfo{person}{Kenji
  Fukumizu}, \bibinfo{person}{Choon~Hui Teo}, \bibinfo{person}{Le Song},
  \bibinfo{person}{Bernhard Sch{\"o}lkopf}, \bibinfo{person}{Alexander~J
  Smola}, {et~al\mbox{.}}} \bibinfo{year}{2007}\natexlab{}.
\newblock \showarticletitle{A kernel statistical test of independence.}. In
  \bibinfo{booktitle}{\emph{NeurIPS}}. \bibinfo{pages}{585--592}.
\newblock


\bibitem[\protect\citeauthoryear{He, Kang, and McAuley}{He
  et~al\mbox{.}}{2017}]%
        {he2017translation}
\bibfield{author}{\bibinfo{person}{Ruining He}, \bibinfo{person}{Wang-Cheng
  Kang}, {and} \bibinfo{person}{Julian McAuley}.}
  \bibinfo{year}{2017}\natexlab{}.
\newblock \showarticletitle{Translation-based recommendation}. In
  \bibinfo{booktitle}{\emph{RecSys}}. \bibinfo{pages}{161--169}.
\newblock


\bibitem[\protect\citeauthoryear{He and McAuley}{He and McAuley}{2016a}]%
        {he2016fusing}
\bibfield{author}{\bibinfo{person}{Ruining He} {and} \bibinfo{person}{Julian
  McAuley}.} \bibinfo{year}{2016}\natexlab{a}.
\newblock \showarticletitle{Fusing similarity models with markov chains for
  sparse sequential recommendation}. In \bibinfo{booktitle}{\emph{ICDM}}.
  \bibinfo{pages}{191--200}.
\newblock


\bibitem[\protect\citeauthoryear{He and McAuley}{He and McAuley}{2016b}]%
        {he2016ups}
\bibfield{author}{\bibinfo{person}{Ruining He} {and} \bibinfo{person}{Julian
  McAuley}.} \bibinfo{year}{2016}\natexlab{b}.
\newblock \showarticletitle{Ups and downs: Modeling the visual evolution of
  fashion trends with one-class collaborative filtering}. In
  \bibinfo{booktitle}{\emph{WWWW}}. \bibinfo{pages}{507--517}.
\newblock


\bibitem[\protect\citeauthoryear{Hidasi, Karatzoglou, Baltrunas, and
  Tikk}{Hidasi et~al\mbox{.}}{2015}]%
        {hidasi2015session}
\bibfield{author}{\bibinfo{person}{Bal{\'a}zs Hidasi},
  \bibinfo{person}{Alexandros Karatzoglou}, \bibinfo{person}{Linas Baltrunas},
  {and} \bibinfo{person}{Domonkos Tikk}.} \bibinfo{year}{2015}\natexlab{}.
\newblock \showarticletitle{Session-based recommendations with recurrent neural
  networks}.
\newblock \bibinfo{journal}{\emph{arXiv preprint arXiv:1511.06939}}
  (\bibinfo{year}{2015}).
\newblock


\bibitem[\protect\citeauthoryear{Hidasi and Tikk}{Hidasi and Tikk}{2016}]%
        {hidasi2016general}
\bibfield{author}{\bibinfo{person}{Bal{\'a}zs Hidasi} {and}
  \bibinfo{person}{Domonkos Tikk}.} \bibinfo{year}{2016}\natexlab{}.
\newblock \showarticletitle{General factorization framework for context-aware
  recommendations}.
\newblock \bibinfo{journal}{\emph{Data Mining and Knowledge Discovery}}
  \bibinfo{volume}{30}, \bibinfo{number}{2} (\bibinfo{year}{2016}),
  \bibinfo{pages}{342--371}.
\newblock


\bibitem[\protect\citeauthoryear{Hinton, Krizhevsky, and Wang}{Hinton
  et~al\mbox{.}}{2011}]%
        {hinton2011transforming}
\bibfield{author}{\bibinfo{person}{Geoffrey~E Hinton}, \bibinfo{person}{Alex
  Krizhevsky}, {and} \bibinfo{person}{Sida~D Wang}.}
  \bibinfo{year}{2011}\natexlab{}.
\newblock \showarticletitle{Transforming auto-encoders}. In
  \bibinfo{booktitle}{\emph{ICANN}}. \bibinfo{pages}{44--51}.
\newblock


\bibitem[\protect\citeauthoryear{Hinton, Sabour, and Frosst}{Hinton
  et~al\mbox{.}}{2018}]%
        {hinton2018matrix}
\bibfield{author}{\bibinfo{person}{Geoffrey~E Hinton}, \bibinfo{person}{Sara
  Sabour}, {and} \bibinfo{person}{Nicholas Frosst}.}
  \bibinfo{year}{2018}\natexlab{}.
\newblock \showarticletitle{Matrix capsules with EM routing}. In
  \bibinfo{booktitle}{\emph{ICLR}}.
\newblock


\bibitem[\protect\citeauthoryear{Jean, Cho, Memisevic, and Bengio}{Jean
  et~al\mbox{.}}{2014}]%
        {jean2014using}
\bibfield{author}{\bibinfo{person}{S{\'e}bastien Jean},
  \bibinfo{person}{Kyunghyun Cho}, \bibinfo{person}{Roland Memisevic}, {and}
  \bibinfo{person}{Yoshua Bengio}.} \bibinfo{year}{2014}\natexlab{}.
\newblock \showarticletitle{On using very large target vocabulary for neural
  machine translation}.
\newblock \bibinfo{journal}{\emph{arXiv preprint arXiv:1412.2007}}
  (\bibinfo{year}{2014}).
\newblock


\bibitem[\protect\citeauthoryear{Johnson, Douze, and J{\'e}gou}{Johnson
  et~al\mbox{.}}{2019}]%
        {johnson2019billion}
\bibfield{author}{\bibinfo{person}{Jeff Johnson}, \bibinfo{person}{Matthijs
  Douze}, {and} \bibinfo{person}{Herv{\'e} J{\'e}gou}.}
  \bibinfo{year}{2019}\natexlab{}.
\newblock \showarticletitle{Billion-scale similarity search with gpus}.
\newblock \bibinfo{journal}{\emph{IEEE Transactions on Big Data}}
  (\bibinfo{year}{2019}).
\newblock


\bibitem[\protect\citeauthoryear{Kang and McAuley}{Kang and McAuley}{2018}]%
        {kang2018self}
\bibfield{author}{\bibinfo{person}{Wang-Cheng Kang} {and}
  \bibinfo{person}{Julian McAuley}.} \bibinfo{year}{2018}\natexlab{}.
\newblock \showarticletitle{Self-attentive sequential recommendation}. In
  \bibinfo{booktitle}{\emph{ICDM}}. IEEE, \bibinfo{pages}{197--206}.
\newblock


\bibitem[\protect\citeauthoryear{Kingma and Ba}{Kingma and Ba}{2014}]%
        {kingma2014adam}
\bibfield{author}{\bibinfo{person}{Diederik~P Kingma} {and}
  \bibinfo{person}{Jimmy Ba}.} \bibinfo{year}{2014}\natexlab{}.
\newblock \showarticletitle{Adam: A method for stochastic optimization}.
\newblock \bibinfo{journal}{\emph{arXiv preprint arXiv:1412.6980}}
  (\bibinfo{year}{2014}).
\newblock


\bibitem[\protect\citeauthoryear{Kuang, Cui, Athey, Xiong, and Li}{Kuang
  et~al\mbox{.}}{2018}]%
        {kuang2018stable}
\bibfield{author}{\bibinfo{person}{Kun Kuang}, \bibinfo{person}{Peng Cui},
  \bibinfo{person}{Susan Athey}, \bibinfo{person}{Ruoxuan Xiong}, {and}
  \bibinfo{person}{Bo Li}.} \bibinfo{year}{2018}\natexlab{}.
\newblock \showarticletitle{Stable prediction across unknown environments}. In
  \bibinfo{booktitle}{\emph{KDD}}. \bibinfo{pages}{1617--1626}.
\newblock


\bibitem[\protect\citeauthoryear{Kuang, Xiong, Cui, Athey, and Li}{Kuang
  et~al\mbox{.}}{2020}]%
        {kuang2020stable}
\bibfield{author}{\bibinfo{person}{Kun Kuang}, \bibinfo{person}{Ruoxuan Xiong},
  \bibinfo{person}{Peng Cui}, \bibinfo{person}{Susan Athey}, {and}
  \bibinfo{person}{Bo Li}.} \bibinfo{year}{2020}\natexlab{}.
\newblock \showarticletitle{Stable prediction with model misspecification and
  agnostic distribution shift}. In \bibinfo{booktitle}{\emph{AAAI}}.
  \bibinfo{pages}{4485--4492}.
\newblock


\bibitem[\protect\citeauthoryear{Kuang, Zhang, Wu, Wu, Zhuang, and Zhang}{Kuang
  et~al\mbox{.}}{2021}]%
        {kuang2021balance}
\bibfield{author}{\bibinfo{person}{Kun Kuang}, \bibinfo{person}{Hengtao Zhang},
  \bibinfo{person}{Runze Wu}, \bibinfo{person}{Fei Wu},
  \bibinfo{person}{Yueting Zhuang}, {and} \bibinfo{person}{Aijun Zhang}.}
  \bibinfo{year}{2021}\natexlab{}.
\newblock \showarticletitle{Balance-Subsampled stable prediction across unknown
  test data}.
\newblock \bibinfo{journal}{\emph{ACM Transactions on Knowledge Discovery from
  Data (TKDD)}} \bibinfo{volume}{16}, \bibinfo{number}{3}
  (\bibinfo{year}{2021}), \bibinfo{pages}{1--21}.
\newblock


\bibitem[\protect\citeauthoryear{Li, Liu, Wu, Xu, Zhao, Huang, Kang, Chen, Li,
  and Lee}{Li et~al\mbox{.}}{2019}]%
        {li2019multi}
\bibfield{author}{\bibinfo{person}{Chao Li}, \bibinfo{person}{Zhiyuan Liu},
  \bibinfo{person}{Mengmeng Wu}, \bibinfo{person}{Yuchi Xu},
  \bibinfo{person}{Huan Zhao}, \bibinfo{person}{Pipei Huang},
  \bibinfo{person}{Guoliang Kang}, \bibinfo{person}{Qiwei Chen},
  \bibinfo{person}{Wei Li}, {and} \bibinfo{person}{Dik~Lun Lee}.}
  \bibinfo{year}{2019}\natexlab{}.
\newblock \showarticletitle{Multi-interest network with dynamic routing for
  recommendation at Tmall}. In \bibinfo{booktitle}{\emph{CIKM}}.
  \bibinfo{pages}{2615--2623}.
\newblock


\bibitem[\protect\citeauthoryear{Li, Wang, and McAuley}{Li
  et~al\mbox{.}}{2020}]%
        {li2020time}
\bibfield{author}{\bibinfo{person}{Jiacheng Li}, \bibinfo{person}{Yujie Wang},
  {and} \bibinfo{person}{Julian McAuley}.} \bibinfo{year}{2020}\natexlab{}.
\newblock \showarticletitle{Time Interval Aware Self-Attention for Sequential
  Recommendation}. In \bibinfo{booktitle}{\emph{WSDM}}.
  \bibinfo{pages}{322--330}.
\newblock


\bibitem[\protect\citeauthoryear{Lin, Feng, Santos, Yu, Xiang, Zhou, and
  Bengio}{Lin et~al\mbox{.}}{2017}]%
        {lin2017structured}
\bibfield{author}{\bibinfo{person}{Zhouhan Lin}, \bibinfo{person}{Minwei Feng},
  \bibinfo{person}{Cicero Nogueira~dos Santos}, \bibinfo{person}{Mo Yu},
  \bibinfo{person}{Bing Xiang}, \bibinfo{person}{Bowen Zhou}, {and}
  \bibinfo{person}{Yoshua Bengio}.} \bibinfo{year}{2017}\natexlab{}.
\newblock \showarticletitle{A structured self-attentive sentence embedding}.
\newblock \bibinfo{journal}{\emph{arXiv preprint arXiv:1703.03130}}
  (\bibinfo{year}{2017}).
\newblock


\bibitem[\protect\citeauthoryear{Liu, Tan, Li, Yang, Zhou, and Hu}{Liu
  et~al\mbox{.}}{2019}]%
        {liu2019single}
\bibfield{author}{\bibinfo{person}{Ninghao Liu}, \bibinfo{person}{Qiaoyu Tan},
  \bibinfo{person}{Yuening Li}, \bibinfo{person}{Hongxia Yang},
  \bibinfo{person}{Jingren Zhou}, {and} \bibinfo{person}{Xia Hu}.}
  \bibinfo{year}{2019}\natexlab{}.
\newblock \showarticletitle{Is a single vector enough? exploring node polysemy
  for network embedding}. In \bibinfo{booktitle}{\emph{KDD}}.
  \bibinfo{pages}{932--940}.
\newblock


\bibitem[\protect\citeauthoryear{Liu, Wu, Wang, and Tan}{Liu
  et~al\mbox{.}}{2016}]%
        {liu2016predicting}
\bibfield{author}{\bibinfo{person}{Qiang Liu}, \bibinfo{person}{Shu Wu},
  \bibinfo{person}{Liang Wang}, {and} \bibinfo{person}{Tieniu Tan}.}
  \bibinfo{year}{2016}\natexlab{}.
\newblock \showarticletitle{Predicting the next Location: A Recurrent Model
  with Spatial and Temporal Contexts}. In \bibinfo{booktitle}{\emph{AAAI}}.
\newblock


\bibitem[\protect\citeauthoryear{Liu, Zeng, Mokhosi, and Zhang}{Liu
  et~al\mbox{.}}{2018}]%
        {liu2018stamp}
\bibfield{author}{\bibinfo{person}{Qiao Liu}, \bibinfo{person}{Yifu Zeng},
  \bibinfo{person}{Refuoe Mokhosi}, {and} \bibinfo{person}{Haibin Zhang}.}
  \bibinfo{year}{2018}\natexlab{}.
\newblock \showarticletitle{STAMP: short-term attention/memory priority model
  for session-based recommendation}. In \bibinfo{booktitle}{\emph{KDD}}.
  \bibinfo{pages}{1831--1839}.
\newblock


\bibitem[\protect\citeauthoryear{Liu, Chen, Li, Yu, McAuley, and Xiong}{Liu
  et~al\mbox{.}}{2021}]%
        {liu2021contrastive}
\bibfield{author}{\bibinfo{person}{Zhiwei Liu}, \bibinfo{person}{Yongjun Chen},
  \bibinfo{person}{Jia Li}, \bibinfo{person}{Philip~S Yu},
  \bibinfo{person}{Julian McAuley}, {and} \bibinfo{person}{Caiming Xiong}.}
  \bibinfo{year}{2021}\natexlab{}.
\newblock \showarticletitle{Contrastive self-supervised sequential
  recommendation with robust augmentation}.
\newblock \bibinfo{journal}{\emph{arXiv preprint arXiv:2108.06479}}
  (\bibinfo{year}{2021}).
\newblock


\bibitem[\protect\citeauthoryear{Luo, Liu, and Liu}{Luo et~al\mbox{.}}{2021}]%
        {luo2021stan}
\bibfield{author}{\bibinfo{person}{Yingtao Luo}, \bibinfo{person}{Qiang Liu},
  {and} \bibinfo{person}{Zhaocheng Liu}.} \bibinfo{year}{2021}\natexlab{}.
\newblock \showarticletitle{STAN: Spatio-Temporal Attention Network for Next
  Location Recommendation}. In \bibinfo{booktitle}{\emph{WWW}}.
  \bibinfo{pages}{2177--2185}.
\newblock


\bibitem[\protect\citeauthoryear{Ma, Zhou, Cui, Yang, and Zhu}{Ma
  et~al\mbox{.}}{2019}]%
        {ma2019learning}
\bibfield{author}{\bibinfo{person}{Jianxin Ma}, \bibinfo{person}{Chang Zhou},
  \bibinfo{person}{Peng Cui}, \bibinfo{person}{Hongxia Yang}, {and}
  \bibinfo{person}{Wenwu Zhu}.} \bibinfo{year}{2019}\natexlab{}.
\newblock \showarticletitle{Learning disentangled representations for
  recommendation}.
\newblock \bibinfo{journal}{\emph{arXiv preprint arXiv:1910.14238}}
  (\bibinfo{year}{2019}).
\newblock


\bibitem[\protect\citeauthoryear{Ma, Zhao, Huang, Wang, Hu, Zhu, and Gai}{Ma
  et~al\mbox{.}}{2018}]%
        {ma2018entire}
\bibfield{author}{\bibinfo{person}{Xiao Ma}, \bibinfo{person}{Liqin Zhao},
  \bibinfo{person}{Guan Huang}, \bibinfo{person}{Zhi Wang},
  \bibinfo{person}{Zelin Hu}, \bibinfo{person}{Xiaoqiang Zhu}, {and}
  \bibinfo{person}{Kun Gai}.} \bibinfo{year}{2018}\natexlab{}.
\newblock \showarticletitle{Entire space multi-task model: An effective
  approach for estimating post-click conversion rate}. In
  \bibinfo{booktitle}{\emph{SIGIR}}. \bibinfo{pages}{1137--1140}.
\newblock


\bibitem[\protect\citeauthoryear{McAuley, Targett, Shi, and Van
  Den~Hengel}{McAuley et~al\mbox{.}}{2015}]%
        {mcauley2015image}
\bibfield{author}{\bibinfo{person}{Julian McAuley},
  \bibinfo{person}{Christopher Targett}, \bibinfo{person}{Qinfeng Shi}, {and}
  \bibinfo{person}{Anton Van Den~Hengel}.} \bibinfo{year}{2015}\natexlab{}.
\newblock \showarticletitle{Image-based recommendations on styles and
  substitutes}. In \bibinfo{booktitle}{\emph{SIGIR}}. \bibinfo{pages}{43--52}.
\newblock


\bibitem[\protect\citeauthoryear{Ni, Li, and McAuley}{Ni et~al\mbox{.}}{2019}]%
        {ni2019justifying}
\bibfield{author}{\bibinfo{person}{Jianmo Ni}, \bibinfo{person}{Jiacheng Li},
  {and} \bibinfo{person}{Julian McAuley}.} \bibinfo{year}{2019}\natexlab{}.
\newblock \showarticletitle{Justifying recommendations using distantly-labeled
  reviews and fine-grained aspects}. In \bibinfo{booktitle}{\emph{EMNLP}}.
  \bibinfo{pages}{188--197}.
\newblock


\bibitem[\protect\citeauthoryear{O'Mahony, Hurley, and Silvestre}{O'Mahony
  et~al\mbox{.}}{2006}]%
        {o2006detecting}
\bibfield{author}{\bibinfo{person}{Michael~P O'Mahony}, \bibinfo{person}{Neil~J
  Hurley}, {and} \bibinfo{person}{Gu{\'e}nol{\'e}~CM Silvestre}.}
  \bibinfo{year}{2006}\natexlab{}.
\newblock \showarticletitle{Detecting noise in recommender system databases}.
  In \bibinfo{booktitle}{\emph{IUI}}. \bibinfo{pages}{109--115}.
\newblock


\bibitem[\protect\citeauthoryear{Rendle, Freudenthaler, and
  Schmidt-Thieme}{Rendle et~al\mbox{.}}{2010}]%
        {rendle2010factorizing}
\bibfield{author}{\bibinfo{person}{Steffen Rendle}, \bibinfo{person}{Christoph
  Freudenthaler}, {and} \bibinfo{person}{Lars Schmidt-Thieme}.}
  \bibinfo{year}{2010}\natexlab{}.
\newblock \showarticletitle{Factorizing personalized markov chains for
  next-basket recommendation}. In \bibinfo{booktitle}{\emph{WWW}}.
  \bibinfo{pages}{811--820}.
\newblock


\bibitem[\protect\citeauthoryear{Sabour, Frosst, and Hinton}{Sabour
  et~al\mbox{.}}{2017}]%
        {sabour2017dynamic}
\bibfield{author}{\bibinfo{person}{Sara Sabour}, \bibinfo{person}{Nicholas
  Frosst}, {and} \bibinfo{person}{Geoffrey~E Hinton}.}
  \bibinfo{year}{2017}\natexlab{}.
\newblock \showarticletitle{Dynamic routing between capsules}.
\newblock \bibinfo{journal}{\emph{arXiv preprint arXiv:1710.09829}}
  (\bibinfo{year}{2017}).
\newblock


\bibitem[\protect\citeauthoryear{Shen, Cui, Zhang, and Kunag}{Shen
  et~al\mbox{.}}{2020}]%
        {shen2020stable}
\bibfield{author}{\bibinfo{person}{Zheyan Shen}, \bibinfo{person}{Peng Cui},
  \bibinfo{person}{Tong Zhang}, {and} \bibinfo{person}{Kun Kunag}.}
  \bibinfo{year}{2020}\natexlab{}.
\newblock \showarticletitle{Stable learning via sample reweighting}. In
  \bibinfo{booktitle}{\emph{AAAI}}. \bibinfo{pages}{5692--5699}.
\newblock


\bibitem[\protect\citeauthoryear{Shen, Liu, He, Zhang, Xu, Yu, and Cui}{Shen
  et~al\mbox{.}}{2021}]%
        {shen2021towards}
\bibfield{author}{\bibinfo{person}{Zheyan Shen}, \bibinfo{person}{Jiashuo Liu},
  \bibinfo{person}{Yue He}, \bibinfo{person}{Xingxuan Zhang},
  \bibinfo{person}{Renzhe Xu}, \bibinfo{person}{Han Yu}, {and}
  \bibinfo{person}{Peng Cui}.} \bibinfo{year}{2021}\natexlab{}.
\newblock \showarticletitle{Towards out-of-distribution generalization: A
  survey}.
\newblock \bibinfo{journal}{\emph{arXiv preprint arXiv:2108.13624}}
  (\bibinfo{year}{2021}).
\newblock


\bibitem[\protect\citeauthoryear{Sun, Liu, Wu, Pei, Lin, Ou, and Jiang}{Sun
  et~al\mbox{.}}{2019}]%
        {sun2019bert4rec}
\bibfield{author}{\bibinfo{person}{Fei Sun}, \bibinfo{person}{Jun Liu},
  \bibinfo{person}{Jian Wu}, \bibinfo{person}{Changhua Pei},
  \bibinfo{person}{Xiao Lin}, \bibinfo{person}{Wenwu Ou}, {and}
  \bibinfo{person}{Peng Jiang}.} \bibinfo{year}{2019}\natexlab{}.
\newblock \showarticletitle{BERT4Rec: Sequential recommendation with
  bidirectional encoder representations from transformer}. In
  \bibinfo{booktitle}{\emph{CIKM}}. \bibinfo{pages}{1441--1450}.
\newblock


\bibitem[\protect\citeauthoryear{Tan, Zhang, Yao, Liu, Zhou, Yang, and Hu}{Tan
  et~al\mbox{.}}{2021}]%
        {tan2021sparse}
\bibfield{author}{\bibinfo{person}{Qiaoyu Tan}, \bibinfo{person}{Jianwei
  Zhang}, \bibinfo{person}{Jiangchao Yao}, \bibinfo{person}{Ninghao Liu},
  \bibinfo{person}{Jingren Zhou}, \bibinfo{person}{Hongxia Yang}, {and}
  \bibinfo{person}{Xia Hu}.} \bibinfo{year}{2021}\natexlab{}.
\newblock \showarticletitle{Sparse-interest network for sequential
  recommendation}. In \bibinfo{booktitle}{\emph{WSDM}}.
  \bibinfo{pages}{598--606}.
\newblock


\bibitem[\protect\citeauthoryear{Tang and Wang}{Tang and Wang}{2018}]%
        {tang2018personalized}
\bibfield{author}{\bibinfo{person}{Jiaxi Tang} {and} \bibinfo{person}{Ke
  Wang}.} \bibinfo{year}{2018}\natexlab{}.
\newblock \showarticletitle{Personalized top-n sequential recommendation via
  convolutional sequence embedding}. In \bibinfo{booktitle}{\emph{WSDM}}.
  \bibinfo{pages}{565--573}.
\newblock


\bibitem[\protect\citeauthoryear{Vaswani, Shazeer, Parmar, Uszkoreit, Jones,
  Gomez, Kaiser, and Polosukhin}{Vaswani et~al\mbox{.}}{2017}]%
        {vaswani2017attention}
\bibfield{author}{\bibinfo{person}{Ashish Vaswani}, \bibinfo{person}{Noam
  Shazeer}, \bibinfo{person}{Niki Parmar}, \bibinfo{person}{Jakob Uszkoreit},
  \bibinfo{person}{Llion Jones}, \bibinfo{person}{Aidan~N Gomez},
  \bibinfo{person}{{\L}ukasz Kaiser}, {and} \bibinfo{person}{Illia
  Polosukhin}.} \bibinfo{year}{2017}\natexlab{}.
\newblock \showarticletitle{Attention is all you need}. In
  \bibinfo{booktitle}{\emph{NeurIPS}}. \bibinfo{pages}{5998--6008}.
\newblock


\bibitem[\protect\citeauthoryear{Wang, Liu, Liu, and Wu}{Wang
  et~al\mbox{.}}{2019}]%
        {wang2019towards}
\bibfield{author}{\bibinfo{person}{Jingyi Wang}, \bibinfo{person}{Qiang Liu},
  \bibinfo{person}{Zhaocheng Liu}, {and} \bibinfo{person}{Shu Wu}.}
  \bibinfo{year}{2019}\natexlab{}.
\newblock \showarticletitle{Towards accurate and interpretable sequential
  prediction: A cnn \& attention-based feature extractor}. In
  \bibinfo{booktitle}{\emph{CIKM}}. \bibinfo{pages}{1703--1712}.
\newblock


\bibitem[\protect\citeauthoryear{Wang, Feng, He, Nie, and Chua}{Wang
  et~al\mbox{.}}{2021}]%
        {wang2021denoising}
\bibfield{author}{\bibinfo{person}{Wenjie Wang}, \bibinfo{person}{Fuli Feng},
  \bibinfo{person}{Xiangnan He}, \bibinfo{person}{Liqiang Nie}, {and}
  \bibinfo{person}{Tat-Seng Chua}.} \bibinfo{year}{2021}\natexlab{}.
\newblock \showarticletitle{Denoising implicit feedback for recommendation}. In
  \bibinfo{booktitle}{\emph{WSDM}}. \bibinfo{pages}{373--381}.
\newblock


\bibitem[\protect\citeauthoryear{Xie, Sun, Liu, Wu, Gao, Ding, and Cui}{Xie
  et~al\mbox{.}}{2020}]%
        {xie2020contrastive}
\bibfield{author}{\bibinfo{person}{Xu Xie}, \bibinfo{person}{Fei Sun},
  \bibinfo{person}{Zhaoyang Liu}, \bibinfo{person}{Shiwen Wu},
  \bibinfo{person}{Jinyang Gao}, \bibinfo{person}{Bolin Ding}, {and}
  \bibinfo{person}{Bin Cui}.} \bibinfo{year}{2020}\natexlab{}.
\newblock \showarticletitle{Contrastive learning for sequential
  recommendation}.
\newblock \bibinfo{journal}{\emph{arXiv preprint arXiv:2010.14395}}
  (\bibinfo{year}{2020}).
\newblock


\bibitem[\protect\citeauthoryear{Xu, Cui, Shen, Zhang, and Zhang}{Xu
  et~al\mbox{.}}{2021}]%
        {xu2021stable}
\bibfield{author}{\bibinfo{person}{Renzhe Xu}, \bibinfo{person}{Peng Cui},
  \bibinfo{person}{Zheyan Shen}, \bibinfo{person}{Xingxuan Zhang}, {and}
  \bibinfo{person}{Tong Zhang}.} \bibinfo{year}{2021}\natexlab{}.
\newblock \showarticletitle{Why Stable Learning Works? A Theory of Covariate
  Shift Generalization}.
\newblock \bibinfo{journal}{\emph{arXiv preprint arXiv:2111.02355}}
  (\bibinfo{year}{2021}).
\newblock


\bibitem[\protect\citeauthoryear{Yu, Liu, Wu, Wang, and Tan}{Yu
  et~al\mbox{.}}{2016}]%
        {yu2016dynamic}
\bibfield{author}{\bibinfo{person}{Feng Yu}, \bibinfo{person}{Qiang Liu},
  \bibinfo{person}{Shu Wu}, \bibinfo{person}{Liang Wang}, {and}
  \bibinfo{person}{Tieniu Tan}.} \bibinfo{year}{2016}\natexlab{}.
\newblock \showarticletitle{A dynamic recurrent model for next basket
  recommendation}. In \bibinfo{booktitle}{\emph{SIGIR}}.
  \bibinfo{pages}{729--732}.
\newblock


\bibitem[\protect\citeauthoryear{Yuan, Hsia, Yang, Zhu, Chang, Dong, and
  Lin}{Yuan et~al\mbox{.}}{2019}]%
        {yuan2019improving}
\bibfield{author}{\bibinfo{person}{Bowen Yuan}, \bibinfo{person}{Jui-Yang
  Hsia}, \bibinfo{person}{Meng-Yuan Yang}, \bibinfo{person}{Hong Zhu},
  \bibinfo{person}{Chih-Yao Chang}, \bibinfo{person}{Zhenhua Dong}, {and}
  \bibinfo{person}{Chih-Jen Lin}.} \bibinfo{year}{2019}\natexlab{}.
\newblock \showarticletitle{Improving ad click prediction by considering
  non-displayed events}. In \bibinfo{booktitle}{\emph{CIKM}}.
  \bibinfo{pages}{329--338}.
\newblock


\bibitem[\protect\citeauthoryear{Zhang, Yao, Zhao, Chua, and Wu}{Zhang
  et~al\mbox{.}}{2021b}]%
        {zhang2021causerec}
\bibfield{author}{\bibinfo{person}{Shengyu Zhang}, \bibinfo{person}{Dong Yao},
  \bibinfo{person}{Zhou Zhao}, \bibinfo{person}{Tat-Seng Chua}, {and}
  \bibinfo{person}{Fei Wu}.} \bibinfo{year}{2021}\natexlab{b}.
\newblock \showarticletitle{Causerec: Counterfactual user sequence synthesis
  for sequential recommendation}. In \bibinfo{booktitle}{\emph{SIGIR}}.
  \bibinfo{pages}{367--377}.
\newblock


\bibitem[\protect\citeauthoryear{Zhang, Cui, Xu, Zhou, He, and Shen}{Zhang
  et~al\mbox{.}}{2021a}]%
        {zhang2021deep}
\bibfield{author}{\bibinfo{person}{Xingxuan Zhang}, \bibinfo{person}{Peng Cui},
  \bibinfo{person}{Renzhe Xu}, \bibinfo{person}{Linjun Zhou},
  \bibinfo{person}{Yue He}, {and} \bibinfo{person}{Zheyan Shen}.}
  \bibinfo{year}{2021}\natexlab{a}.
\newblock \showarticletitle{Deep Stable Learning for Out-Of-Distribution
  Generalization}. In \bibinfo{booktitle}{\emph{CVPR}}.
  \bibinfo{pages}{5372--5382}.
\newblock


\bibitem[\protect\citeauthoryear{Zhou, Ma, Zhang, Zhou, and Yang}{Zhou
  et~al\mbox{.}}{2021}]%
        {zhou2021contrastive}
\bibfield{author}{\bibinfo{person}{Chang Zhou}, \bibinfo{person}{Jianxin Ma},
  \bibinfo{person}{Jianwei Zhang}, \bibinfo{person}{Jingren Zhou}, {and}
  \bibinfo{person}{Hongxia Yang}.} \bibinfo{year}{2021}\natexlab{}.
\newblock \showarticletitle{Contrastive learning for debiased candidate
  generation in large-scale recommender systems}. In
  \bibinfo{booktitle}{\emph{KDD}}. \bibinfo{pages}{3985--3995}.
\newblock


\bibitem[\protect\citeauthoryear{Zhou, Wang, Zhao, Zhu, Wang, Zhang, Wang, and
  Wen}{Zhou et~al\mbox{.}}{2020}]%
        {zhou2020s3}
\bibfield{author}{\bibinfo{person}{Kun Zhou}, \bibinfo{person}{Hui Wang},
  \bibinfo{person}{Wayne~Xin Zhao}, \bibinfo{person}{Yutao Zhu},
  \bibinfo{person}{Sirui Wang}, \bibinfo{person}{Fuzheng Zhang},
  \bibinfo{person}{Zhongyuan Wang}, {and} \bibinfo{person}{Ji-Rong Wen}.}
  \bibinfo{year}{2020}\natexlab{}.
\newblock \showarticletitle{S3-rec: Self-supervised learning for sequential
  recommendation with mutual information maximization}. In
  \bibinfo{booktitle}{\emph{CIKM}}. \bibinfo{pages}{1893--1902}.
\newblock


\end{thebibliography}

\end{document}